\documentclass[12pt]{article}

\usepackage{graphics,epsfig}
\usepackage{amsbsy}
\usepackage{amsfonts}
\usepackage{amsmath}
\usepackage{graphicx}

\def\Xint#1{\mathchoice
{\XXint\displaystyle\textstyle{#1}}%
{\XXint\textstyle\scriptstyle{#1}}%
{\XXint\scriptstyle\scriptscriptstyle{#1}}%
{\XXint\scriptscriptstyle\scriptscriptstyle{#1}}%
\!\int}
\def\XXint#1#2#3{{\setbox0=\hbox{$#1{#2#3}{\int}$ }
\vcenter{\hbox{$#2#3$ }}\kern-.6\wd0}}

\def\dashint{\Xint-}

\def\res#1#2{\mbox{Res}\left[#1,#2\right]}
\def\ppint{\dashint}

\def\id{\protect{{1 \kern-.28em {\rm l}}}}

\newcommand{\beq}{\begin{equation}}
\newcommand{\eeq}{\end{equation}}
\newcommand{\beqr}{\begin{displaymath}}
\newcommand{\eeqr}{\end{displaymath}}
\newcommand{\beqa}{\begin{eqnarray}}
\newcommand{\eeqa}{\end{eqnarray}}
\newcommand{\beqar}{\begin{eqnarray*}}
\newcommand{\eeqar}{\end{eqnarray*}}

\def\k{\kappa}

\def\p{{\partial}}
\def\nn{\nonumber}

\newcommand{\adss}[2]{\mbox{$AdS_{#1}\times {S}^{#2}$}}

\def\dalemb#1#2{{\vbox{\hrule height .#2pt
        \hbox{\vrule width.#2pt height#1pt \kern#1pt
                \vrule width.#2pt}
        \hrule height.#2pt}}}

\def\half{\frac{1}{2}}
\let\a=\alpha \let\b=\beta \let\g=\gamma \let\d=\delta \let\e=\epsilon
\let\z=\zeta  \let\th=\theta  \let\k=\kappa
\let\l=\lambda \let\m=\mu  \let\x=\xi \let\p=\pi 
\let\s=\sigma \let\t=\tau   \let\c=\chi 
 \let\vep=\varepsilon
\let\w=\omega      \let\G=\Gamma \let\D=\Delta \let\Th=\Theta \let\L=\Lambda
 \let\P=\Pi \let\S=\Sigma  
\let\C=\Chi \let\W=\Omega
\let\la=\label \let\ci=\cite 
  
\def\nn{\nonumber} \def\bd{\begin{document}} \def\ed{\end{document}}
\def\ds{\documentstyle} \let\fr=\frac \let\bl=\bigl \let\br=\bigr
\let\Br=\Bigr \let\Bl=\Bigl
\let\bm=\bibitem
\let\na=\nabla
\def\tU{{\widetilde U}}
\let\pa=\partial \let\ov=\overline
\def\ie{{\it i.e.\ }}
\newcommand{\be}{\begin{equation}}
\newcommand{\ee}{\end{equation}}
\def\ba{\begin{array}}
\def\ea{\end{array}}
\def\ft#1#2{{\textstyle{{\scriptstyle #1}\over {\scriptstyle #2}}}}
\def\fft#1#2{{#1 \over #2}}
\def\F#1#2{{ F_{#1}^{(#2)} }}
\def\cF#1#2{{ {\cal F}_{#1}^{(#2)} }}

\def\={\, =\, }
\def\+{\, +\, }
\def\-{\, -\, }

\def\R{{\bf R}}
\def\sst#1{{\scriptscriptstyle #1}}
\def\oneone{\rlap 1\mkern4mu{\rm l}}
\def\e7{E_{7(+7)}}
\def\td{\tilde}
\def\wtd{\widetilde}
\def\im{{\rm i}}
\newcommand{\ho}[1]{$\, ^{#1}$}
\newcommand{\hoch}[1]{$\, ^{#1}$}
\newcommand{\bea}{\begin{eqnarray}}
\newcommand{\eea}{\end{eqnarray}}
\newcommand{\ra}{\rightarrow}
\newcommand{\lra}{\longrightarrow}
\newcommand{\Lra}{\Leftrightarrow}
\newcommand{\ap}{\alpha^\prime}
\newcommand{\bp}{\tilde \beta^\prime}
\newcommand{\cB}{{\cal B}}
\newcommand{\cO}{{\cal O}}
\newcommand{\vecx}{\vec{x}}
\newcommand{\vecy}{\vec{y}}
\newcommand{\vecp}{\vec{p}}
\newcommand{\vecq}{\vec{q}}
\newcommand{\tr}{{\rm tr} }
\newcommand{\Tr}{{\rm Tr} }

\newcommand{\cL}{{\cal L}}
\newcommand{\cA}{{\cal A}}
\newcommand{\cD}{{\cal D}}
\def\sst#1{{\scriptscriptstyle #1}}

\def\ve{\varepsilon}
\def\vf{\varphi}
\def\F{\Phi}
\def\wg{\wedge}

\def \foot {\footnote}
\def \bi{\bibitem}

\def \tr {{\rm tr}}
\def \ha {{1 \over 2}}
\def \td {\tilde}
\def \ci{\cite}
\def \N {{\mathcal N}}
\def \ww {\Omega}
\def \const {{\rm const}}
\def \ss {\sum_{i=1}^3 }
\def \t {\tau}
\def\S{{\mathcal S} }
\def \nn {\nu}
\def \XX {{\rm X}}

\def \lra {\leftrightarrow}
\def \vom {{\bar \omega}}
\def \E {{\mathcal  E}} \def \J {{\mathcal  J}}
\def \YY {{\rm Y}}

\def \d {\del}
\def \rJ {{J}}
\def \sms {sigma models\ }
\def \sm {sigma model\ }
\def \L {\Lambda}
\def \gl {\ell}
\def \tr {{\rm tr\ }}
\def\z{\zeta}
\def\zi{\zeta_1}
\def\zii{\zeta_2}
\def\K{\mbox{K}}
\def\eE{\mbox{E}}   \def \vt {\vartheta}
\def \vr {\varrho}
\def \wup {w}

\def\dg{\dagger}
\def\a{\alpha}
\def\b{\beta}
\def\e{\varepsilon}
\def\p{\phi}
\def\ap{\alpha^\prime}
\def\I{{\cal I}}

\def\R{{\bf R}}
\def\Z{{\bf Z}}
\def\C{{\bf C}}
\def\P{{\bf P}}
\def\xb{{\bar X}}
\def\Tr{{\rm  Tr}}
\def\tr{{\rm  tr}}

\def \del{\partial}
\def \a {\alpha}
\def \aa {{\a'}}
\def\g{\gamma}
\def\s{\sigma}
\def\z{\zeta}
\def\zi{\zeta_1}
\def\zii{\zeta_2}
\def\ov{\over}

\def\I{{\cal I}}
\def\J{{\mathcal J}}
\def \ok {{1\ov \k}}
\def\LL{{\mathcal L }}
\def \jL {{J}}
\def \om {\omega}
\def \cL {{\mathcal L}} \def \cH {{\mathcal H}}
\def\E{{\mathcal E}}
\def\w{\omega}
\def\b{\beta}
\def\l{\lambda}
\def\eps{\epsilon}
\def\vep{\varepsilon}
\def \De {{\mathcal D}}

 \def \cV {{\cal V}}
\def  \Jt {  {J}_{\rm tot}    }

\def \k {\kappa}
\def\foot{\footnote}
\def \four{{\textstyle {1\ov 4}}}
 \def \third { \textstyle {1\ov 3
}}
\def\det{\hbox{det}}
\def \ci {\cite}

\def \foot {\footnote}
\def \bi{\bibitem}

\def \tr {{\rm tr}}
\def \ha {{1 \over 2}}
\def \tid {\tilde}
\def \vv {{\rm v}}
\def \tl {{\tilde \l}}
\def \XX {{\rm X}}
\def \ta {{\tilde \a}}
\def \fo { {1\ov 4}}
\def \ep {\epsilon}
\def \inti {{\int^{2\pi}_0 {d \sigma \ov 2 \pi}}}

\def \d {\partial}
\def \K {{\rm S}}
\def \el {\ell}
\def \Tr {{\rm Tr}}
\def \P {\Phi}
\def \l  {\lambda}
\def \tl {{\tilde \l}}
\def \bl {{\tilde \l}}
\def \const {{\rm const}}
\def \V {v}

\def \bv {v^*}
\def \vv {{\rm v}}
\def \LL {{\mathcal L}}
\newcommand{\PV}[1]{P_{\!\!_{V_{#1}}}}

\def \bL {\ell}
\def \M {{\mathcal M}}
\def \N {{\mathcal N}}
\def \S {{\rm S}}
\def \vn {\vec n}
\def \tl {\td \l}
\def \td {\tilde}
\def \Prod {\Pi}
\def \O {{\mathcal O}}
\def \Q {{\rm  Q}}
\def \D {\Delta}
\def \N {{\mathcal N}}
\def\tN{{\tilde N}}

\def \m {\mu}
\def \vs {\vec \s}
\def \ie {i.e.}

\def \cD {{\cal D}}

\def  \le  {\l_{\rm eff}}

\def \rS {{\rm S}}
\def\as{{\a}}
\newcommand{\bra}[1]{\mbox{$\langle #1 |$}}
\newcommand{\ket}[1]{\mbox{$| #1 \rangle$}}

\newcommand{\Gg}{G}

\newcommand{\auth}{AUTHORS}

\def\thb{\bar{\theta}}
\def\Thb{\bar{\Theta}}
\def\barp{\bar{p}}
\def\barq{\bar{q}}
\def\barc{\bar{c}}
\def\bard{\bar{d}}
\def\e{\epsilon}

\def \bi{\bibitem}
\def \la {\label}

\def \l {\lambda}
\def\foot{\footnote}
\def \tl  {{\tilde \l}}
\def \sql {{\sqrt \l}}
\def \adss {$AdS_5 \times S^5$\ }
\newcommand{\rf}[1]{(\ref{#1})}
\def \ov {\over}

\def\th{\theta}
\def\Th{\Theta}
\def\vth{\vartheta}

\def\vth{\vartheta}

\def\ra{\rightarrow}
\def\N{{\cal N}}
\def\F{{\cal F}}
\def\cc{\circ}
\def\eqv{\equiv}

\def\ni{\noindent}
\def \ha{{1\ov 2}}
\def \bw {{\rm w}}

\def\r{{\rm r}}

\def \cT {{\cal T}}
\def \no {\nonumber}

\def \J {\mathcal{J}}
\def \del {\partial}

\def \bps {{\bar \psi}}
\def \sqbl {\sqrt{\bar \lambda}}

\def\dF{\dot{F}}
\def\dG{\dot{G}}
\def\df{\dot{f}}
\def \E {{\cal E}}

\def \S {{\cal S}}
\def \J {{\cal J}}

\def\ms{\mathcal{S}}
\def\mj{\mathcal{J}}
\def\soj{\fr{\ms}{\mj}}
\def \R {{\bf R}}
\def \om {\omega}
\def \tH {\widetilde H}
\def \bE {\bar E}
\def \x {{\cal X}}

\def \hV {{\hat V}}

 \def \bb {\bar \beta}
\def \W {{\cal E}}

\def \bi{\bibitem}
\def \la {\label}

\def \l {\lambda}
\def\foot{\footnote}
\def \tl  {{\tilde \l}}
\def \sql {{\sqrt \l}}
\def \sqtl {{\sqrt {\tilde \l}}}

\def \HH {{\rm E}}
\def \cS {{\cal S}}
\def \cL {{\cal L}}

\def \adss {$AdS_5 \times S^5$\ }

\def \D {\Delta}
\def \thet {\theta}
 \def \t {\tau}
 \def \p {\phi}
 \def \r {\rho}
 \def \rN {{\rm N}}
 \def\tw{{\tilde w}}
 \def\hJ{{J}}
 \def\hw{{w}}
 \def\hl{{\lambda}}
 \def\hth{{\theta}}
 \def\NN{{\cal N}}
 \def \bv {{ \bar w}}
\def \vn {{\vec n}}
\newcommand{\sfrac}[2]{{\textstyle\frac{#1}{#2}}}
\def \bl {{ \bar \lambda}}
\def \bp {{\bar p}}
\def \bu {{\bar u}}
\def \sha {\sfrac{1}{2}}
\def \w {\omega}
\def \ov {\over}
\def \vl { \vec \ell}
\def \varpi {{\rm w}}
\def \OO {{\cal O}}
\def \bG {\bar \G}

\def \c {\gamma}

\def \ss {{\rm s}}

\def \ve {\varepsilon}
\def \pa{\partial}
\def \I {{\cal I}}
\def \LL {{\cal L}}
\def \ep {\epsilon}
\def \R {{\rm R}}
\def \tilt {{\tilde t}}

\def\pic #1#2{\hbox{\lower#1pt\hbox{~\mbox{\epsfxsize=20truemm \epsffile{#2}}}}}

\def\pic #1#2#3{\hbox{\lower#1pt\hbox{~\mbox{\includegraphics[scale=#3]{#2}}}}}

\def \bt {\bar\theta}
\def \te {\theta}

\def \cc {{\rm f}}
\def \d {\delta}

\def \cL {{\cal L}}
\def \S  {{\cal S}}
\def \pp {{q}}
\def \vt {\vartheta}
\def \mm {{\cal  \ell}}
\def \Z {{\cal Z}}
\def \pa {\partial}
\def \C {{\cal C}}
\def \be {\bea}
\def \ee {\eea}
\def \c {\gamma}  \def \d {\delta}
\def \eps {\epsilon}

\def \bp {\begin{pmatrix}}  \def \ep {\end{pmatrix}}

 \def \T {{\cal T}}

\def \bp {\begin{pmatrix}}  \def \epm {\end{pmatrix}}
\def \ha {{\textstyle{1 \ov 2}}}

\begin{document}
\overfullrule=0pt
\parskip=2pt
\parindent=12pt
\headheight=0in \headsep=0in \topmargin=0in \oddsidemargin=0in

\vspace{ -3cm} \thispagestyle{empty} \vspace{-1cm}
\begin{flushright} UUITP-14/09
\end{flushright}
\begin{center}
 \vspace{2cm}
{\Large\bf
 Spiky strings in the SL(2) Bethe Ansatz
 }

 \vspace{.5cm} {
  L. Freyhult$^{a,}$\footnote{lisa.freyhult@physics.uu.se},
  M. Kruczenski$^{b,}$\footnote{markru@purdue.edu}
 and A.
 Tirziu$^{b,}$\footnote{atirziu@purdue.edu}\\
 \vskip 0.3cm

{\em
$^{a}$   Department of Physcis and Astronomy, Uppsala University,\\
 P.O. Box 803, S-75108, Uppsala, Sweden  \\
\vskip 0.08cm
$^{b}$Department of Physics, Purdue  University,\\
W. Lafayette, IN 47907-2036, USA.\\ }}

\end{center}

 \begin{abstract}
We study spiky strings in the context of the SL(2) Bethe ansatz equations. We find an asymmetric distribution of Bethe roots along one cut
that determines the all loop anomalous dimension at leading and subleading orders in a large S expansion. At leading order in strong coupling (large $\lambda$) we obtain that the energy of such states is given, in terms of the spin $S$ and the number of spikes $n$ by $E-S=\frac{n \sqrt{\lambda}}{2 \pi}\left(\ln \frac{4\pi S}{ \sqrt{\lambda}}+\ln \frac{4}{n}\sin\frac{\pi}{n}-1\right)+ \mathcal{O}\left(\frac{\ln S}{S}\right).$
This result matches perfectly the same expansion obtained from the known spiky string  classical solution.

We then discuss a two cut spiky string Bethe root distribution at one-loop in the SL(2) Bethe ansatz. In this case we find a limit where $n\rightarrow \infty$, keeping
$\frac{E+S}{n^2}$, $\frac{E-S}{n}$, $\frac{J}{n}$ fixed. This is the one loop version of a limit previously considered in the context of the string classical solutions in $AdS_5 \times S^5$. In that case it was related to a string solution in the $AdS$ $pp$-wave background.

\end{abstract}
\newpage



\renewcommand{\theequation}{1.\arabic{equation}}
 \setcounter{equation}{0}

\setcounter{equation}{0} \setcounter{footnote}{0}
\setcounter{section}{0}

\section{Introduction}

Remarkable progress was achieved in the understanding of the AdS/CFT correspondence by studying the $SL(2)$ sector of the theory.
Much of this progress relates to twist two gauge theory operators of the type
$\tr(\Phi D_{+}^S \Phi)$. On the string side they are described
by the folded string solution \cite{gkp} as can also be verified by an alternative computation in terms of the cusp anomaly \cite{cusp}.
In the field theory
a proposed all loop Bethe ansatz \cite{bes} can be used to compute their anomalous dimension at all loops in the planar approximation.
Computations of quantum corrections to the folded string solution \cite{ft1} at leading order in large $S$ as well as its generalization
to $AdS_3 \times S^1$ \cite{ftt,rtt,rt2} were recently used to successfully check such all loop asymptotic Bethe ansatz  \cite{bbks, bkk,grom, frs}.
The all loop asymptotic Bethe ansatz computation was then extended to the subleading orders in large $S$ expansion \cite{frs,fz,fgr} and precise matching with the
corresponding string theory computations \cite{bftt} was found.

 Beyond twist two, the $SL(2)$ sector contains operators of higher twist which have also been investigated. For example operators of the type
\beq\la{opa}
\cO = \tr\left( D_+^{\frac{S}{n}}\Phi\ D_+^{\frac{S}{n}}\Phi\ldots D_+^{\frac{S}{n}}\Phi\  \right)
\eeq
are described,
on the string side, by the spiky string solutions \cite{spiky}. There has been some recent progress in this area. In \cite{Jevicki}
the solutions
were shown to correspond to multi-soliton solutions of a generalized sinh-Gordon model. This allowed the construction of new, more general, solutions
where the spikes move with respect to each other opening up the possibility to study a whole new class of operators. The relation with field theory was examined
in detail in \cite{Dorey1} where the elliptic curves associated with the classical solution where analyzed and a map was proposed
to a similar structure emerging from the study of the field theory operators.

 However, with the current methods, full understanding of these operators from the field theory point of view requires the construction of the Bethe
ansatz solution that describes them. It is the goal of this paper to fill this gap by providing a proposal to describe the spiky string solution in
the all loop Bethe ansatz.

It was shown in \ci{bgk,kz} and extended in \cite{Dorey1} that such operators can be described by a spin chain  with a  number of sites $n$  being the
same as the number of  spikes. At leading order in large $S$ the spiky string solution touches the boundary of $AdS_3$. It was shown \cite{kt} that
at this order the spiky string solution can be mapped to the folded string solution by a conformal transformation, and thus the all loop energy at
leading order in large $S$ is given by
\begin{equation}
E- S= \frac{n}{2}f(\lambda) \ln S + \mathcal{O}(S^{0},n)  \label{yat}
\end{equation}
where $f(\lambda)$ is a universal scaling function known as the cusp anomaly. In this paper we obtain this result from the field theory side by an all loop Bethe ansatz computation.

 In a semiclassical analysis in string theory the spiky string solution was extended to $AdS_3 \times S^1$ in \cite{iktt}. It turns out that, in the limit
of large number of spikes $n \rightarrow\infty$, there is an interesting scaling limit in which $\frac{E+S}{n^2}$, $\frac{E-S}{n}$ and $\frac{J}{n}$ remain
fixed. It was shown in \cite{iktt} that the solution in this limit is related to a solution in the $AdS$ $pp$-wave background introduced in \cite{kt}.
In this paper, by using the found Bethe ansatz solution, it is shown that the same $pp$-wave like limit can be taken at weak coupling.

 To construct the Bethe ansatz solution, an important input is a set of integers $n_k$ which may be interpreted as the bosonic
mode numbers of the waves propagating in the spin chain. Different such sets of numbers give rise to different solutions of the Bethe ansatz equations.
 The folded string solution in $AdS_3$ reduces \cite{tt} to the folded string solution in flat space in the limit $\frac{S}{\sqrt{\lambda}}\ll 1$.
In flat space we can quantize the string exactly in terms of right and left moving waves propagating on the string. It is easily seen that the folded
string has the same number of right and left moving excitations, i.e.
$n_L=n_R$, with the same wave numbers $k_L=k_R=1$. Extending
this idea to $AdS_3$
for all values of spin $S$ and coupling $\lambda$, suggests that in a Bethe ansatz computation the folded string solution is described by two sets of
modes, one with  $n_k=-1$ and the other with  $n_k=1$. These numbers were used then as the input in the one loop SL(2) Bethe ansatz equations for
two cuts (corresponding to $n_k=\pm1$), and a symmetric Bethe root distribution was found to describe the folded string solution \cite{bfst}.
The same bosonic quantum numbers were used later \cite{es, bes} to find the corresponding solution to the all loop Bethe ansatz. In the large $S$
limit and finite $J$ the two cuts in fact merge into one cut with a discontinuous wave-number distribution since we still have $n_k=\pm1$.

Following the same procedure, we look at the spiky string solution in flat space, find the left and right wave numbers and use them as an input to
find the corresponding Bethe root equation that we then proceed to solve.  The spiky string solution was already discussed in \cite{spiky}. It turns
out that the bosonic quantum numbers that should be used in the Bethe ansatz equations are $n_L=-1$ and $n_R=n-1$. Using this input we find an asymmetric
distribution of Bethe roots spread over one cut along the real axis at one loop in weak coupling and leading order in the large $S$ expansion. We then
extend the computation to all loops and further obtain the all loop result also for the subleading order in large $S$. More precisely, we obtain that
the energy of the spiky string at all loops can be written as
\begin{equation}
E-S=n+f(g)\frac{n}{2}\left(\ln S+\gamma_E+\ln\left(\frac{2}{n}\sin\frac{\pi}{n}\right)\right)+\frac{n}{2}B_2(g)+\mathcal{O}(\frac{\ln S}{S},n),
 \quad \quad g^2= \frac{\lambda}{16 \pi^2}  \label{jio}
\end{equation}
where $B_2(g)$ denotes the virtual scaling function of twist 2 operators obtained in \cite{fz}. The leading large $S$ term, \ie\
$E-S\simeq f(g)\frac{n}{2} \ln S$, matches the string theory result (\ref{yat}) at the same order.

 If we further expand the all loop result (\ref{jio}) for large coupling  we obtain
\begin{eqnarray}
E-S&=&\frac{n \sqrt{\lambda}}{2 \pi}\left(\ln \frac{4 \pi S}{\sqrt{\lambda}}+\ln \left(\frac{4}{n}\sin\frac{\pi}{n}\right)-1\right)\\
&+&\frac{n}{2}\left(1+\frac{6\ln2}{\pi}-\frac{3\ln 2}{\pi}\ln\left(\frac{4}{n}\sin\frac{\pi}{n}\right)-\frac{3\ln 2}{\pi}\ln\frac{4\pi S}{\sqrt{\lambda}}\right)+\mathcal{O}(\frac{1}{\sqrt{\lambda}}, \frac{\ln S}{S})\nonumber  \label{wju}
\end{eqnarray}
Remarkably, the first line precisely matches the result known from expanding the classical string solution \cite{bftt}, that is, not only at order
$\ln S$ but also $S^0$. The second line is a prediction for the quantum corrections to the classical solution which will be interesting to check explicitly.
Let us observe that, since the asymptotic Bethe ansatz reproduces the correct strong coupling result to order $S^0$, wrapping effects play no role to this order in $S$.

 Having understood the solution with one cut we proceed to consider a two cut solution to the one-loop Bethe ansatz equations.
Again we use the ``spiky string'' bosonic quantum numbers for the integers $n_k$, \ie\ one cut has $n_k=-1$ and the other $n_k=n-1$ where $n$
is the number of spikes. The solution has four real parameters $d<c<a<b$ which represent the position of the two cuts given by the segments $[d,c]$
and $[b,a]$. The solution obtained allows us to compute the energy $E$, spin $S$, R-charge $J$ and number of spikes $n$ in terms of these parameters.
Although there are four parameters it turns out that, for the solution to exist, a consistency condition has to be satisfied which implies that only
three are independent. Equivalently this means that we can, implicitly, express the energy as a function of the other physical quantities $E=E(S,J,n)$.
As a check, in the limit $c,b\rightarrow 0$, namely when the cuts merge, we recover the results of the previous one cut solution.

 More interestingly, in the limit $d \rightarrow - \infty$ a $pp$-wave type scaling is obtained, i.e. $n\rightarrow \infty$, with $\frac{E+S}{n^2}$,
$\frac{E-S}{n}$ and $\frac{J}{n}$ remaining fixed. In this limit the equations simplify since one parameter is removed. To be able to compare with
the string calculation of \cite{iktt} it would be necessary to extend this results to all loops and then consider the strong coupling limit. This should be
an important test of the all-loop Bethe ansatz since these solutions contain a lot of structure and a highly non-trivial dependence of the
energy with the other parameters.

Another direction one can pursue is to consider the Bethe ansatz description of the more general solutions described in \cite{Jevicki}. In fact at this
point it might be more interesting to find a direct mapping between the classical action of the string and the classical action of the spin chain
(at strong coupling) in a similar way as in \cite{spinchain}. Mapping the actions would determine a mapping between the configurations of the string and
those of the spin chain. In particular this would allow to understand how to see the spikes and their motion from the spin chain/field theory point of
view. In the approach we follow here this cannot be seen since we map  a semiclassical or coherent state on the string side to an eigenstate of the
Hamiltonian (Bethe ansatz). A more detailed map
would have to be constructed between coherent states on both sides.

The paper is organized as follows. In section 2 we review the spiky string solution in flat space which leads us to the correct mode numbers to be used
in the Bethe ansatz. In section 3 we present in detail the construction of the one-loop Bethe root distribution for a one cut solution at leading order in
large $S$. In section 4 we extend the computation to all loops using the all loop asymptotic Bethe ansatz. Section 5 deals with the extension of the all
loop solution at leading order in large $S$ to the subleading order $\mathcal{O}(S^0)$. In section 6 we analyze the spiky string solution with two cuts
using the one loop Bethe ansatz equations. We then take two limits: first a limit when the two cuts collide recovering the one cut solution, and second,
we consider the $pp$-wave type limit. In appendix A we present some details about the solution of the Bethe ansatz equations which we used throughout
the paper. In Appendix B we present a review of the $\mathcal{O}(S^0)$ corrections to the ground state, which corresponds to the folded spinning
string solution.

\renewcommand{\theequation}{2.\arabic{equation}}
 \setcounter{equation}{0}

\setcounter{equation}{0} \setcounter{footnote}{0}

\section{Spiky strings in flat space}
\label{flat space}

 As with the folded string, it is convenient to start by studying solutions in flat space. In that case we can quantize
the theory exactly and  it turns out that the solutions are just a superpositions of a left and a right moving waves:
\beqa
 x &=& A\, \cos\left((n-1)\ \sigma_+\right) + A\, (n-1)\, \cos\left(\sigma_-\right) \\
 y &=& A\, \sin\left((n-1)\ \sigma_+\right) + A\, (n-1)\, \sin\left(\sigma_-\right) \\
 t &=& 2\, A\, (n-1)\, \tau = A\, (n-1)\,(\sigma_++\sigma_-)
\label{flatsol}
\eeqa
 where $\sigma_+ = \tau+\sigma$, $\sigma_-=\tau-\sigma$, $n$ is the number of spikes and $A$ is constant that
determines the size of the string. Here, ($\tau,\sigma$) parameterize the world-sheet of a string which is moving
in a Minkowski space with metric:
\beq
 ds^2 = -dt^2 + dx^2+dy^2
\eeq
 The solutions are periodic in $\sigma$ with period $2\pi$ and satisfy the equations of motion in conformal gauge, $
 (\partial_\tau^2 - \partial_\sigma^2) X^{\mu} = 0 $, as well as the constraints $ (\partial_+ X)^2 = (\partial_-X)^2 =0$.
 Quantum mechanically the state has $n_R = A^2 (n-1)^2$ right moving excitations of wave number $k_R=1$
and $n_L = A^2 (n-1)$ left moving excitations with wave number $k_L=n-1$ (satisfying the level matching
condition $n_Rk_R =n_L k_L$).

 All excitations carry one unit of angular momentum and therefore the total angular momentum and energy are given by
\beq
S = n_L+n_R = A^2 n (n-1) , \ \ E=\sqrt{2(n_Lk_L+n_Rk_R)}= 2A (n-1), \ \ E = 2\sqrt{\frac{n-1}{n}S}
\eeq
which agrees with a classical computation. For $n=2$ we recover the standard Regge
trajectory $E=\sqrt{2S}$ and for $n>2$ we get a Regge trajectory of modified slope.

\begin{figure}
\epsfig{file=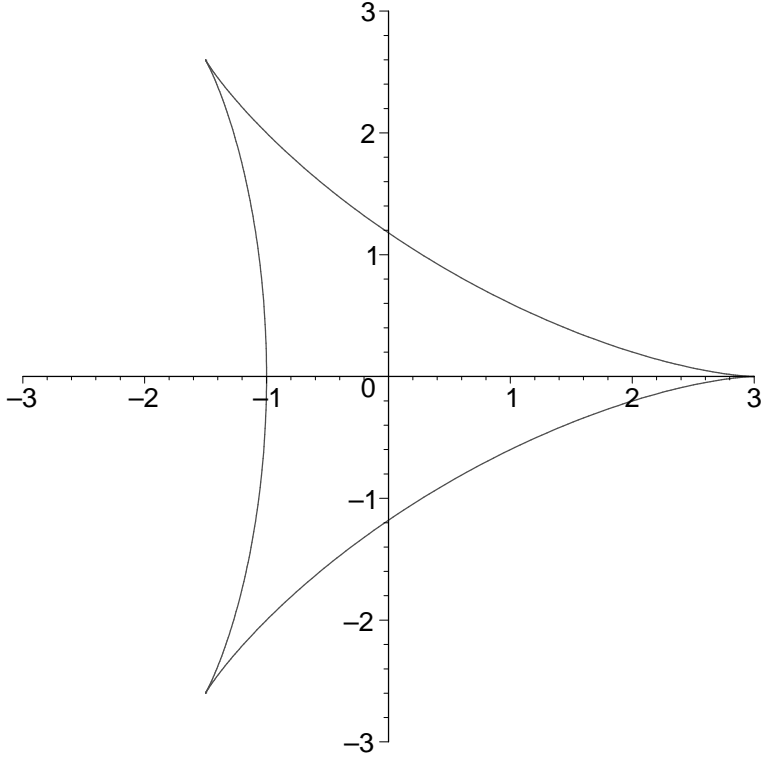, height=10cm}
\caption{Rotating string with 3 spikes in flat space. We plot $(x,y)$ in eq.(\ref{flatsol})
parametrically as a function of $\sigma=0\rightarrow2\pi$ for $n=3$, $A=1$ and $\tau=0$.
 }
\label{fig:3flat}
\end{figure}

\begin{figure}
\epsfig{file=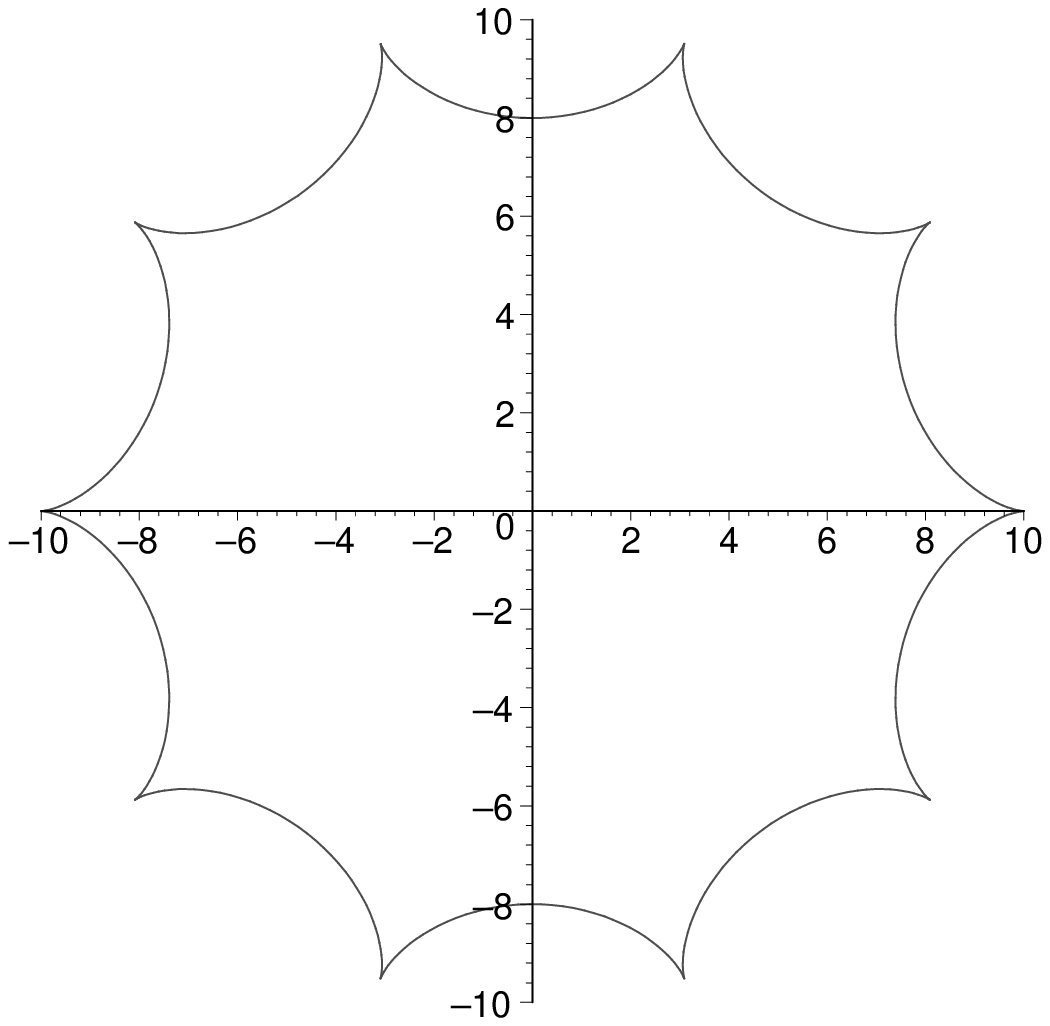, height=10cm}
\caption{Rotating string with 10 spikes in flat space. We plot $(x,y)$ in eq.(\ref{flatsol})
parametrically as a function of $\sigma=0\rightarrow2\pi$ for $n=10$, $A=1$ and $\tau=0$.}
\label{fig:10flat}
\end{figure}

At fixed time, the shape of the string for different values of $n$ takes the form
depicted in figs.(\ref{fig:3flat}) and (\ref{fig:10flat}). It can be seen that the string has $n$
spikes or cusps and,  analyzing the time dependence, that it rotates rigidly in such a way that the end points of the spikes
move at the speed of light.  More details on these solutions can be found in \cite{spiky}. Here, we will just need the fact that
the spiky string is obtained by superposing left and right moving waves of different momentum. This suggests that when extending
to all loops we should take a distributions of Bethe roots such that the wave number can take the values $n_w=-1$
or $n_w=n-1$ where $n$ is the number of spikes.

\renewcommand{\theequation}{3.\arabic{equation}}
 \setcounter{equation}{0}

\setcounter{equation}{0} \setcounter{footnote}{0}

\section{Spiky strings in the Bethe Ansatz:  $1$-cut solution}
\label{AdS}

We follow the same prescription used in \cite{es} to obtain the $\ln S$ dependence from the Bethe ansatz at large $S$.
In the SL(2) sector the Bethe roots are on the real axis and, for large $S$, they accumulate on various segments or cuts.
In the case of the folded string with $J=2$ and $S \gg 1$, there is only one cut and the root distribution is symmetric\footnote{In the case when $S \sim J \gg 1$ there are two cuts. This situation was analyzed in \cite{bfst} and will be studied later in this paper.}. In this case, namely for the lowest twist, there is only one state whereas
for $J > 2$ there is more then one state. Nevertheless, for the lowest energy state at higher twist, the root distribution is again symmetric as discussed
in \cite{es}. In this section we consider excited energy states which describe the $n$-spike solution in the large $S$ limit for a fixed finite $J$.

Following the discussion of the previous section, in the case of the spiky strings at leading order in large $S$, we assume to have one cut but
with an asymmetric root distribution. The mode numbers of each excitation will be taken to be $n_w=-1$ or $n_w=n-1$ where $n$ is the number of spikes.

Let us start with the one loop SL(2) spin chain. The one loop Bethe ansatz equations corresponding to a $XXX_{-\frac{1}{2}}$ nearest neighbor spin chain can be written as
\begin{equation}
\bigg(\frac{u_k+ \frac{i}{2}}{u_k-\frac{i}{2}}\bigg)^J=\prod_{j \neq k}^S \frac{u_k - u_j-i}{u_k - u_j +i}, \quad \quad k=1,2,...,S  \label{qop}
\end{equation}
Taking the logarithm of (\ref{qop}), and further expanding in $\frac{1}{u}$
as appropriate for large $S$ gives \cite{es}
\begin{equation}
\frac{J}{u_k}=2 \pi n_k -2 \sum_{j=1,j \neq k}^S \frac{1}{u_k-u_j} \label{ba}
\end{equation}
where $u_k$ are the Bethe roots and $J$ is the length of the corresponding spin chain. The total momentum should vanish due to the cyclicity of the trace.
This condition reads
\begin{equation}
\prod_{j=1}^S \frac{u_j + \frac{i}{2}}{u_j-\frac{i}{2}}=1  \label{mo}
\end{equation}
The one loop energy is given by
\begin{equation}
\frac{E-S-J}{2g^2} = \sum_{k=1}^S \frac{1}{u_k^2+\frac{1}{4}}
\end{equation}
At this point one should solve eq.(\ref{ba}) for the real numbers $u_k$ subject to the condition (\ref{mo}). However, when the number of roots $S$ is very
large, the roots accumulate in cuts on the real axis and one can conveniently approximate the problem by defining a density of roots.
Thus we introduce the root distribution function $\rho_0(u)=\sum_k \delta(u- u_k)$. Furthermore, for large $S$, the cut
grows as $S$ and we can make the approximation that $u\gg 1$.
Therefore, at leading order in the large $S$ expansion (\ref{ba}), we need to solve the equations
\begin{eqnarray}
\ppint d u \frac{\rho_0(u)}{u'-u} &=& \pi n_w(u')  \label{ht} \\
\int d u \rho_0(u) &=& S \label{ht1}\\
\ppint d u \frac{\rho_0(u)}{u}  &=&0, \label{ht2}
\end{eqnarray}
which are, the Bethe equation, the normalization of the density, and the zero momentum condition. The function $\rho_0(u)$ has support on a finite interval that needs to be determined as part of the solution.

To solve these equations we use the method described in the appendix. We first define the cut, namely the interval where $\rho_0$ is non-vanishing,
to extend from $d<x<a$ with $d<0$ and $a>0$. The function $n_w(u)$ is defined as
\begin{eqnarray}
n_w(u)=\left\{
       \begin{array}{ll}
         -1, & d < u < 0 \\
         n-1, & 0 < u < a
       \end{array}
     \right. \label{alp}
\end{eqnarray}
In figure 3 we plot such functions in terms of a rescaled $\bar{u}=\frac{u}{S}$.
\begin{figure}
\epsfig{file=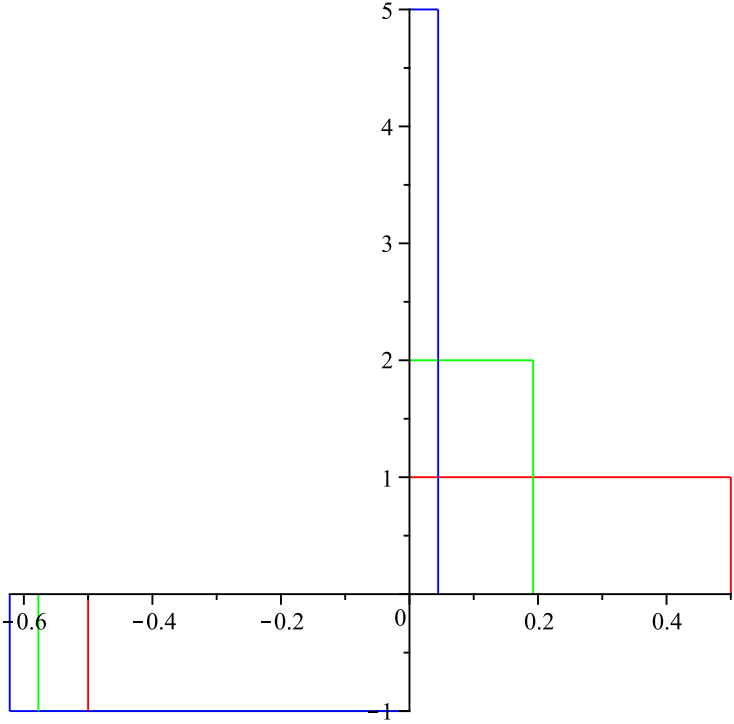, height=10cm}
\caption{Wave numbers $n_w(\bar{u})=n_w(\frac{u}{S})$ for n=2,3,6. By analogy with flat space, we take the modes with $\bar{u}<0$ to have $n_w=-1$
and those with $\bar{u}>0$ to have $n_w=n-1$.}
\label{fig3}
\end{figure}
We then introduce a function  $F(w)=\sqrt{w-a}\sqrt{w-d}$ which has a cut on the real axis extending precisely from $d$ to $a$. Furthermore, right above (below)
the cut it is purely imaginary with positive (negative) imaginary part. Using this
function we define the resolvent as
\beq\label{defresolvent}
 \Gg(w) = \frac{1}{\pi} \int_d^a \frac{F(w)}{F(u+i0^+)} \frac{n_w(u)}{u-w} du
\eeq
which, following the appendix A, determines $\rho_0(u)$ to be
\beq
 \rho_0(u) = \frac{1}{\pi} \ppint_d^a \left|\frac{F(u)}{F(u')}\right|\frac{n_w(u')}{u'-u} du' \label{rhores}
\eeq
Besides, we need the behavior of $\Gg(w)$ at infinity which is readily seen to be
\beq
 \Gg(w) \simeq \Gg_0 + \Gg_1 \frac{1}{w} + \ldots , \ \ \ w\rightarrow\infty
\eeq
with
\beqa
 \Gg_0 &=& -\frac{1}{\pi} \int_d^a \frac{n_w(u)}{F(u+i0^+)} du =i\frac{n-2}{2} + \frac{in}{\pi}\arcsin\left(\frac{a+d}{a-d}\right) \label{H0}\\
 \Gg_1 &=& -\frac{a+d}{2} \Gg_0 -\frac{1}{\pi}  \int_d^a \frac{u n_w(u)}{F(u+i0^+)} du  \label{H1}
\eeqa
 As explained in the appendix we need to satisfy the consistency condition
\beq
 \res{\frac{\Gg(w)}{w-u}}{\infty} = -\Gg_0 =0 \label{conscond0}
\eeq
for $\rho_0(u)$ to solve eq.(\ref{ht}). On the other hand, equation (\ref{ht1}) gives
\beq
 S= \int_d^a \rho_0(u) du = \half \oint_C \Gg(w) dw = i \pi \res{\Gg(w)}{\infty} = -i\pi \Gg_1 \label{Seq}
\eeq
and eq.(\ref{ht2}) is
\beq
 0=\ppint_d^a \frac{\rho_0(u)}{u} du = \half \oint_C \frac{\Gg(w)}{w} dw = i \pi \res{\frac{\Gg(w)}{w}}{\infty}  = i\pi \Gg_0
\eeq
which is satisfied if (\ref{conscond0}) is. In the previous equations we used a contour $C$ that encircles the cut in a clockwise direction.
 Using \eqref{H0} we then find
\beq
 -\frac{a}{d} = \tan^2\left(\frac{\pi}{2n}\right) \label{conscond}
\eeq
Using \eqref{H1} and \eqref{Seq} gives
\beq
S = n \sqrt{-ad} \label{Sres}
\eeq
Before computing the energy let us compute the generating function $\Gg$
\beqa\label{resolventcomplex}
G(w)= \frac{F(w)}{i \pi} \int_d^a d u \frac{n_w(u)}{(u-w)\sqrt{(a-u)(u-d)}}=
-i \pm \frac{n}{\pi}\ln \frac{\sqrt{a (w-d)}+\sqrt{-d (a-w)}}{\sqrt{-d (a-w)}-\sqrt{a(w-d)}}
\eeqa
where $\pm$ refer to $w$ having positive or negative imaginary part. $w$ takes values in the complex plane with a cut along the real axis from $d$ to $a$.

 We now need to compute the energy which, within the Bethe ansatz is given by
\beqa
\frac{E - S-J}{2g^2} &=& \int_d^a \frac{\rho_0(u)}{u^2+\frac{1}{4}} du = \half \oint_C \frac{\Gg(w)}{w^2+\frac{1}{4}} dw \\
      &=& i\pi\left\{\res{\frac{\Gg(w)}{w^2+\frac{1}{4}}}{\frac{i}{2}} + \res{\frac{G(w)}{w^2+\frac{1}{4}}}{-\frac{i}{2}}\right\}
          =\pi\left(\Gg(\frac{i}{2})-\Gg(-\frac{i}{2})\right)
\eeqa
 Using the definition of $\Gg(w)$ and through an explicit computation we find
\beqa
\frac{E - S-J}{2g^2}&=& 2\pi \mathrm{Re}G(\frac{i}{2}) = 2 \mbox{Im} \int \frac{F(\frac{i}{2})}{|F(u)|} \frac{n_w(u)}{u-\frac{i}{2}} d u \\
    &=& n \ln\left| \frac{\left(\sqrt{a(\frac{i}{2}-d)}+\sqrt{(a-\frac{i}{2})(-d)}\right)^2}{-\frac{i}{2}(a-d)}\right|
\eeqa
 For fixed $n$ and large $S$ we see from eqs.(\ref{conscond}) and (\ref{Sres}) that $a\sim d\sim S\gg1$ so we can approximate the last equation as
\beq
\frac{E - S-J}{2g^2} \cong n\ln\left(-\frac{8ad}{a-d}\right) \label{Eres}, \quad \quad \quad S \gg 1
\eeq
Therefore, through eqs.(\ref{Sres}), (\ref{Eres}),  we have solved the problem by writing $E$ and $S$ in terms of two parameters
$a$ and $d$ related by the condition (\ref{conscond}). Explicitly we can write
\beqa
 a &=& \frac{S}{n} \tan\frac{\pi}{2n} \\
 d &=& -\frac{S}{n} \cot\frac{\pi}{2n} \\
\frac{E-S-J}{2g^2} &=&  2n \ln \left(\frac{4S}{n}\sin\frac{\pi}{n}\right)
\eeqa
Notice that up to now we have not needed the actual density $\rho_0(u)$. However, in the following section it will be needed for extending
the result to all loops and subleading terms in the large $S$ expansion. Using eq.(\ref{rhores}) it can be written as
\beqa\label{oneloopdensity}
\rho_0(u) &=& \frac{n}{\pi} \ln\left|\frac{\sqrt{a(u-d)}+\sqrt{-d(a-u)}}{\sqrt{a(u-d)}-\sqrt{-d(a-u)}}\right| \\
&=&\left\{
     \begin{array}{ll}
         \frac{2 n}{\pi} \mbox{arctanh} \frac{\sqrt{-d}\sqrt{a-u}}{\sqrt{a}\sqrt{u-d}}, \quad \quad & a \geq u > 0 \\
       \frac{2 n}{\pi} \mbox{arctanh} \frac{\sqrt{a}\sqrt{u-d}}{\sqrt{-d}\sqrt{a-u}}, \quad \quad & d \leq u < 0
     \end{array}
   \right.
 \label{qma1}
\eeqa
Plots of the root distribution are shown in Figure 4.
\begin{figure}
\epsfig{file=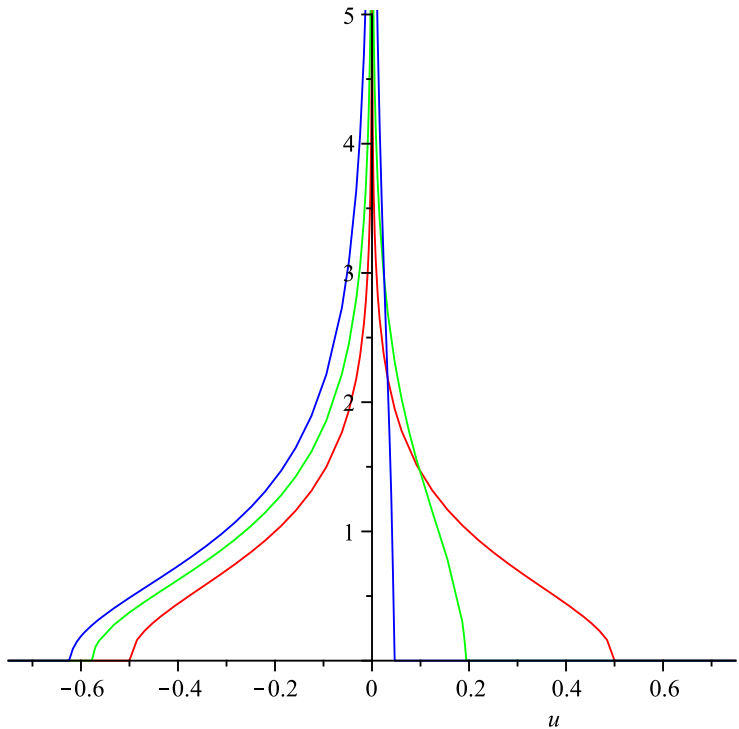, height=10cm}
\caption{Root density $\rho_0(\bar{u})=\rho_0(\frac{u}{S})$ for $n=2$, $n=3$ and $n=6$. The larger the $n$ the more skewed the distribution becomes.}
\label{rho_asym}
\end{figure}
It is clear that $S$ can be thought as determining the scale of the variable $u$. For that reason it is convenient to define a rescaled variable
$\bar{u}=\frac{u}{S}$. Defining new parameters
\beq
\tilde{a} = \frac{1}{n} \cot\frac{\pi}{n}, \ \ \ \tilde{b} = \frac{1}{n\sin\frac{\pi}{n}}
\eeq
such that $\frac{d}{S}=-(\tilde{a}+\tilde{b})$, $\frac{a}{S}=\tilde{b}-\tilde{a}$ we find another convenient way to write the density:
\begin{eqnarray}
\bar{\rho}_0(\bar{u})= \frac{n}{2 \pi}\ln \frac{\tilde{b}^2 - (\bar{u}+\tilde{a}) \tilde{a} + \sqrt{\tilde{b}^2-(\bar{u}+\tilde{a})^2}\sqrt{\tilde{b}^2-\tilde{a}^2}}{\tilde{b}^2 - (\bar{u}+\tilde{a}) \tilde{a} - \sqrt{\tilde{b}^2-(\bar{u}+\tilde{a})^2}\sqrt{\tilde{b}^2-\tilde{a}^2}}
 \label{qma}
\end{eqnarray}
Summarizing, the result from this section is that the energy of the spiky string at one-loop is given by
\begin{equation}
\frac{E - S}{2g^2} = 2 n \ln S + \mathcal{O}(S^0)
\end{equation}
When $n=2$ this indeed recovers the twist two result in \cite{es}. The leading order expression is proportional to $n$ similarly as at strong coupling.

\renewcommand{\theequation}{4.\arabic{equation}}
 \setcounter{equation}{0}

\setcounter{equation}{0} \setcounter{footnote}{0}

\section{Spiky strings in the all loop Bethe ansatz}

After understanding the root distribution at one-loop we now proceed to find a solution to the all-loop Bethe equations. We will follow the procedure
used in \cite{es}.
The asymptotic all loop Bethe equations in the SL(2) sector read
\begin{eqnarray}
2J\arctan(2u_k)+iJ\ln\frac{1+g^2/(x_k^-)^2}{1+g^2/(x_k^+)^2}&=&2\pi \tilde{n}_k-2\sum_{j=-S/2 \atop {j\neq0}}^{S/2}\arctan(u_k-u_j)\\
&+&2i\sum_{j=-S/2 \atop j\neq0}^{S/2}\ln\frac{1-g^2/x_k^+x_j^-}{1-g^2/x_k^-x_j^+}-2\sum_{j=-S/2\atop j\neq 0}^{S/2}\theta(u_k,u_j)\nonumber,
\end{eqnarray}
where $\theta(u,v)$ denotes the dressing phase and $x^\pm(u)$ are defined through $u\pm\tfrac{i}{2}=x^\pm(u)+\tfrac{g^2}{x^\pm(u)}$.
For the ground state the "fermionic" mode numbers are given by \cite{es}
\begin{equation}
\tilde{n}_k=k'+\frac{J-2}{2}\epsilon(k),\quad k'=\pm\frac{1}{2},\,\pm\frac{3}{2},\dots,\pm\frac{S-1}{2},
\end{equation}
where $k'=k-\epsilon(k)/2$.

For the highest excited state there is no gap in the mode numbers \cite{bgk}
\begin{equation}
\tilde{n}_k=k',\quad k'=\pm\frac{1}{2},\,\pm\frac{3}{2},\dots,\pm\frac{S-1}{2}.
\end{equation}
In the limit of large $S$ we define the density of roots as
\begin{equation}
\rho(u)=S\frac{d(k'/S)}{du}
\end{equation}
and the all loop equations can then be written as
\begin{eqnarray}
&&\nonumber\frac{J}{u^2+1/4}+i J\frac{d}{du}\ln\frac{1+g^2/(x^-(u))^2}{1+g^2/(x^+(u))^2}=2\pi\rho(u)-2\int_a^d \frac{\rho(u')}{(u-u')^2+1}\\
&+&\int_a^d du'\rho(u')\frac{d}{du}\left(2i\ln\frac{1-g^2/x^+(u)x^-(u')}{1-g^2/x^+(u')x^-(u)}-2\theta(u,u')\right).
\end{eqnarray}
At one-loop this reduces to
\begin{equation}\label{one-loopsubleading}
\frac{J}{u^2+1/4}=2\pi\rho_0(u)-2\int_{d}^{a}du'\frac{\rho_0(u')}{(u-u')^2+1}.
\end{equation}
Rescaling the roots, $u=S\bar{u}$, we find in the large $S$ limit
\begin{eqnarray}
\frac{2J}{S}\pi\delta(u)=2\pi\bar{\rho}_0(\bar{u})-2\int_{-\tilde{b}-\tilde{a}}^{\tilde{b}-\tilde{a}}d\bar{u}'\bar{\rho}_0(\bar{u})\left(\pi\delta(\bar{u}-\bar{u}')+\frac{1}{S}\mathcal{P}\frac{1}{(\bar{u}-\bar{u}')^2}\right)
\end{eqnarray}
and the one-loop equation reduces to
\begin{equation}
0=\pi J\delta(\bar{u})+\ppint_{-\tilde{b}-\tilde{a}}^{\tilde{b}-\tilde{a}}d\bar{u}'\frac{\bar{\rho}_0(\bar{u}')}{(\bar{u}-\bar{u}')^2}.
\end{equation}
For the highest excited state $J$ is equal to the number of cusps, $n$, and we recognize the above equation as the derivative of the one-loop equation.

 To simplify the presentation of the all-loop computation we will omit the dressing phase and only restore its contribution in the end. We can now proceed with the all loop equation by splitting off the one-loop piece of the density,
 $\rho(u)=\rho_0(u)+2g^2\tilde{\sigma}(u)$,
\begin{eqnarray}
\nonumber0&=&2\pi\tilde{\sigma}(u)-2\int_{-\infty}^\infty\frac{\tilde{\sigma}(u')}{(u-u')^2+1}\\
&+&\nonumber\frac{1}{2g^2}\int_{d}^{a}du'\rho_0(u')\frac{d}{du}\left(2i\ln\frac{1-g^2/x^+(u)x^-(u')}{1-g^2/x^+(u')x^-(u)}\right)\\
&+&\int_{-\infty}^{\infty}du'\tilde{\sigma}(u')\frac{d}{du}\left(2i\ln\frac{1-g^2/x^+(u)x^-(u')}{1-g^2/x^+(u')x^-(u)}\right).
\end{eqnarray}
In the above we have extended the integral boundaries to $\pm\infty$ in the terms containing the higher loop density, $\tilde{\sigma}(u)$.
For the term containing the one-loop density we find, in the large $S$ limit,
\begin{eqnarray}
&&\nonumber\frac{1}{2g^2}\int_{d}^{a}du'\rho_0(u')\frac{d}{du}\left(2i\ln\frac{1-g^2/x^+(u)x^-(u')}{1-g^2/x^+(u')x^-(u)}\right)\\
&&\nonumber=-i\int_{d}^{a}du'\rho_0(u')\frac{d}{du}\left(\frac{1}{x^+(u)x^-(u')}-\frac{1}{x^-(u)x^+(u')}\right)\\
&&=\pi\frac{d}{du}\left(\frac{G\left(\tfrac{i}{2}\right)}{x^+(u)}-\frac{G\left(-\tfrac{i}{2}\right)}{x^-(u)}\right)+\dots,
\end{eqnarray}
Using \eqref{resolventcomplex}
we find
\begin{equation}
G\left(\pm\tfrac{i}{2}\right)=\pm \frac{n}{\pi}\ln S+\dots
\end{equation}
and
\begin{eqnarray}
\nonumber0&=&2\pi\tilde{\sigma}(u)-2\int_{-\infty}^\infty\frac{\sigma(u')}{(u-u')^2+1}\\
\nonumber&+&n\ln S\frac{d}{du}\left(\frac{1}{x^+(u)}+\frac{1}{x^-(u)}\right)\\
&+&\int_{-\infty}^{\infty}du'\tilde{\sigma}(u')\frac{d}{du}\left(2i\ln\frac{1-g^2/x^+(u)x^-(u')}{1-g^2/x^+(u')x^-(u)}\right).
\end{eqnarray}
After Fourier transformation\footnote{We use the convention $\hat{f}(t)=\int_{-\infty}^\infty dw\,e^{-iwt}f(w)$. Details on the Fourier transforms used here can be found in \cite{es}.} and a redefinition of the density,
\begin{equation}
\hat{\sigma}(t)=-2ne^{t/2}\hat{\tilde{\sigma}}(t)\ln S,
\end{equation}
the result is
\begin{equation}
\hat{\sigma}(t)=\frac{t}{e^t-1}\left(K(2gt,0)-4g^2\int_0^\infty dt'K(2gt,2gt')\hat{\sigma}(t')\right)
\end{equation}
with the kernel defined by
\begin{eqnarray}\label{kernelwithoutdressing}
&&K(t,t')=K_0(t,t')+K_1(t,t')\\
&&K_0(t,t')=\frac{2}{tt'}\sum_{n=1}^\infty (2n-1)J_{2n-1}(t)J_{2n-1}(t')\\
&&K_1(t,t')=\frac{2}{tt'}\sum_{n=1}^\infty 2nJ_{2n}(t)J_{2n}(t').
\end{eqnarray}
Including the dressing phase in the computation replaces the kernel above with
\begin{eqnarray}\label{kernelwithdressing}
K(t,t')=K_0(t,t')+K_1(t,t')+K_d(t,t')
\end{eqnarray}
where
\begin{equation}
K_d(t,t')=8g^2\int_0^\infty dt'' K_1(t,2gt'')\frac{t''}{e^{t''}-1}K_0(2gt'',t').
\end{equation}
Further we find that the energy is given by
\begin{equation}
E-S=8g^2\,n\hat{\sigma}(0)\ln S+\mathcal{O}(S^0).
\end{equation}
which gives
\begin{equation}
E-S=\frac{n}{2}f(g)\ln S+\mathcal{O}(S^0)
\end{equation}
which matches the string theory result at leading order in large $S$ \cite{kt}.

\renewcommand{\theequation}{5.\arabic{equation}}
 \setcounter{equation}{0}

\setcounter{equation}{0} \setcounter{footnote}{0}

\section{Subleading corrections for the highest excited state}
The computation of the first subleading corrections for the highest excited state follow the same logic as the corresponding computation for the ground state but uses an asymmetric root density. The subleading corrections for the ground state were computed in \cite{frs,fz} using a method different than what will be convenient here. We will here use the same method as in \cite{es,bes} but also include subleading corrections, for convenience we include the corresponding computation for the ground state in Appendix B.
\subsection{One loop}
The one-loop equation is given in \eqref{one-loopsubleading}, where $J$ should be set to $n$ for the highest excited state.
An approximate solution to this equation is given by \eqref{oneloopdensity} which we expand for large $S$
\begin{eqnarray}
\tilde{\rho}_0(u)
&=&\frac{n}{\pi}\ln S-\frac{n}{2\pi }\ln(u^2)+\frac{n}{\pi }\ln\left(\frac{2}{n}\sin\frac{\pi}{n}\right)+\mathcal{O}(1/S).
\end{eqnarray}
The above density captures only the leading order in the large spin expansion correctly. To get the subleading corrections of order $S^0$ we split off the above solution from the density,
\begin{equation}
\rho_0(u)=\tilde{\rho}_0(u)+r(u),
\end{equation}
and solve for $r(u)$. We are now allowed to extend the limits of integration to $\pm\infty$.
\begin{eqnarray}
\nonumber0&=&2\pi r(u)+2n\ln S-n\ln(u^2)+2n\ln\left(\frac{2}{n}\sin\frac{\pi}{n}\right)-\frac{n}{u^2+1/4}\\
&&-2\int_{-\infty}^\infty dv\left(\frac{n}{\pi}\ln S-\frac{n}{2\pi }\ln(u^2)+\frac{n}{\pi }\ln\left(\frac{2}{n}\sin\frac{\pi}{n}\right)\right)\frac{1}{(u-v)^2+1}\nonumber\\
&-&2\int_{-\infty}^\infty dv\frac{r(v)}{(u-v)^2+1}.
\end{eqnarray}
The equation can be solved by Fourier transformation
\begin{equation}
\hat{r}(t)=n\left(\frac{e^{-|t|/2}}{1-e^{-|t|}}-\frac{1}{|t|}\right).
\end{equation}
Computing the energy including the first two orders in the large $S$ expansion we find
\begin{eqnarray}\label{one-loopEnergy}
\nonumber \frac{E-S-n}{2g^2}&=&\int_{d}^{a}\frac{\tilde{\rho}_0(u)}{u^2+1/4}+\int_{-\infty}^\infty du\frac{r(u)}{u^2+1/4}\\
\nonumber&=&2n\ln S+2n\ln\left(\frac{4}{n}\sin\frac{\pi}{n}\right)+n\int_{-\infty}^\infty dt\left(\frac{e^{-|t|}}{1-e^{-|t|}}-\frac{e^{-|t|/2}}{|t|}\right)\\
&=&2n\left(\ln S+\gamma_E+\ln\left(\frac{2}{n}\sin\frac{\pi}{n}\right)\right).
\end{eqnarray}
%
\subsection{All loops}
The all loop equation can be written as
\begin{eqnarray}
&&n\frac{d}{du}\ln\frac{x(i/2+u)}{x(i/2-u)}=2\pi\rho(u)-\int_{d}^{a}\frac{\rho(u')}{(u-u')^2+1}\nonumber\\
&&+\int_{d}^{a}du'\rho(u')\frac{d}{du}\left(2i\ln\frac{1-g^2/x^+(u)x^-(u')}{1-g^2/x^+(u')x^-(u)}-2\theta(u,u')\right).
\end{eqnarray}
Again we will omit the dressing phase, only to restore it in the end.
Splitting off the one-loop density
\begin{equation}
\rho(u)=\rho_0(u)+2g^2\tilde{\sigma}(u),
\end{equation}
where $\rho_0(u)$ satisfies \eqref{one-loopsubleading},
we find
\begin{eqnarray}\label{all-loopBE}
&&\frac{n}{2g^2}\left(\frac{d}{du}\ln\frac{x(i/2+u)}{x(i/2-u)}-\frac{1}{u^2+1/4}\right)=2\pi\tilde{\sigma}(u)-2\int_{-\infty}^\infty\frac{\tilde{\sigma}(u')}{(u-u')^2+1}+I(u)\nonumber\\
&&+\int_{-\infty}^{\infty}du'\tilde{\sigma}(u')\frac{d}{du}\left(2i\ln\frac{1-g^2/x^+(u)x^-(u')}{1-g^2/x^+(u')x^-(u)}\right)
\end{eqnarray}
where
\begin{eqnarray}
I(u)=\frac{1}{2g^2}\int_{d}^{a}du'\rho_0(u')\frac{d}{du}\left(2i\ln\frac{1-g^2/x^+(u)x^-(u')}{1-g^2/x^+(u')x^-(u)}\right).
\end{eqnarray}
Note that the limits of integration in the integrals containing the higher loop density can be extended to $\pm\infty$ to this level of approximation.
To obtain the all loop energy,
\begin{equation}\label{Eall}
\frac{E-S-n}{2g^2}=\int_d^a\rho(u)\left(\frac{i}{x^+(u)}-\frac{i}{x^-(u)}\right),
\end{equation}
we can follow a similar strategy as above. We split off the one-loop density and then extend the limits of integration in \eqref{Eall} to $\pm\infty$. It is then clear that we only need the combination $\tilde{\sigma}(u)+\tilde{\sigma}(-u)$ to obtain the energy. To write the integral equation for the symmetric combination $\tilde{\sigma}(u)+\tilde{\sigma}(-u)$ we need the combination $I(u)+I(-u)$ which we can split into two parts
\begin{eqnarray}\label{I(u)}
\nonumber I(u)+I(-u)&=&\pi \left(G(\tfrac{i}{2})-G(-\tfrac{i}{2})\right)\frac{d}{du}\left(\frac{1}{x^+(u)}-\frac{1}{x^-(u)}\right)\\
&+&\int_{-\infty}^\infty du'\rho_0(u')\left[\frac{1}{2g^2}\frac{d}{du}\left(2i\ln\frac{1-g^2/x^+(u)x^-(u')}{1-g^2/x^+(u')x^-(u)}\right)\right.\nonumber\\
&-&\left.\frac{1}{u'^2+1/4}\frac{d}{du}\left(\frac{1}{x^+(u)}-\frac{1}{x^-(u)}\right)\right]
\end{eqnarray}

Note that in the integral in \eqref{I(u)} we have extended the integration limits to $\pm \infty$.

Fourier transformation of the equation \eqref{all-loopBE} gives
\begin{eqnarray}
 &&\nonumber \frac{n}{2g^2}2\pi e^{-t/2}\left(J_0(2gt)-1\right)=\pi\left(\hat{\tilde{\sigma}}(t)+\hat{\tilde{\sigma}}(-t)\right)(1-e^{-t})+\frac{1}{2}\left(\hat{I}(t)+\hat{I}(-t)\right)\\
&&+4\pi g^2te^{-t/2}\int_{0}^\infty dt'\left(\hat{\tilde{\sigma}}(t')+\hat{\tilde{\sigma}}(-t')\right)e^{-t'/2}K(2gt,2gt'),
\end{eqnarray}
where the kernel is given by \eqref{kernelwithoutdressing}.
For $\hat{I}(t)$ we find, to first two orders in the expansion,
\begin{eqnarray}
\nonumber \frac{1}{2}\left(\hat{I}(t)+\hat{I}(-t)\right)&=&8\pi n  \left(\ln S+\gamma_E+\ln\left(\frac{2}{n}\sin\frac{\pi}{n}\right)\right)t e^{-t/2}K(2gt,0)\\
&+&8\pi t e^{-t/2}\int_0^\infty dt' \frac{n}{e^{t'}-1}\left(K(2gt,2gt')-K(2gt,0)\right).
\end{eqnarray}
Using the result for the one-loop energy \eqref{one-loopEnergy}
and introducing the density, $\hat{\sigma}(t)$
\begin{equation}
\hat{\sigma}(t)=-\frac{1}{8}e^{-t/2}\left(\hat{\tilde{\sigma}}(t)+\hat{\tilde{\sigma}}(-t)\right)
\end{equation}
we hence find for the all-loop equation
\begin{eqnarray}\label{eq: complete density}
\hat{\sigma}(t)&=&\frac{t}{e^t-1} \Big[ K(2gt,0)\frac{n}{2}\left(\ln S+\gamma_E+\ln\left(\frac{2}{n}\sin\frac{\pi}{n}\right)\right) \nonumber-\frac{n}{8g^2 t}(J_0(2gt) - 1)\\
\nonumber& +&\frac{1}{2}\int_{0}^{\infty} dt'  \frac{n}{e^{t'}-1}\left(K(2gt,2gt')-K(2gt,0)\right) \\
&-& 4g^2 \int_{0}^{\infty} dt' K(2gt,2gt')\hat{\sigma}(t') \Big] \,.
\end{eqnarray}
Taking the dressing function into account replaces the kernel by \eqref{kernelwithdressing}.
The string energy can now be written as
\begin{equation}
E-S=n+16g^2\hat{\sigma}(0).
\end{equation}
We note that the integral equation is
very similar to the equation in \cite{fz} and the solution can consequently be written as
\begin{equation}
E-S=n+f(g)\frac{n}{2}\left(\ln S+\gamma_E+\ln\left(\frac{2}{n}\sin\frac{\pi}{n}\right)\right)+\frac{n}{2}B_2(g)+\dots
\end{equation}
where $B_2(g)$ denotes the virtual scaling function of twist 2 operators. Using the first orders in the expansions at strong coupling \cite{fz},
\begin{eqnarray}
&&\hspace{-1cm}f(g)=4g-\frac{3\ln 2}{\pi}+\mathcal{O}(1/g)\\
&&\hspace{-1cm}B_2(g)=\left(-\gamma_E-\ln g\right)f(g)-4g(1-\ln 2)-\left(1-\frac{6\ln 2}{\pi}+\frac{3(\ln 2)^2}{\pi}\right)+\mathcal{O}(1/g)
\end{eqnarray}
we find
\begin{eqnarray}\label{strongcouplingres}
\nonumber E-S&=&2ng\left(\ln\frac{S}{g}+\ln\left(\frac{4}{n}\sin\frac{\pi}{n}\right)-1\right)\\
&+&\frac{n}{2}\left(1+\frac{6\ln2}{\pi}-\frac{3\ln 2}{\pi}\ln\left(\frac{4}{n}\sin\frac{\pi}{n}\right)-\frac{3\ln2}{\pi}\ln\frac{S}{g}\right)+\mathcal{O}(1/g) \label{shj}
\end{eqnarray}
We see that the leading strong coupling result is in agreement with the known string theory result \cite{bftt}. For $n=2$ we recover the result for the folded string \cite{fz}, this result is in full agreement with the one-loop computation from string theory \cite{bftt, Gromov}. For arbitrary $n$ \eqref{strongcouplingres} is expected to be in agreement with the one-loop result computed from the sigma model \cite{Gromov:2007ky}. Nevertheless,  as we mentioned already, it would be interesting to compare the result in the second line in (\ref{shj}) with an explicit one-loop string computation of the $\mathcal{O}(S^0)$ term.


\renewcommand{\theequation}{6.\arabic{equation}}
 \setcounter{equation}{0}

\setcounter{equation}{0} \setcounter{footnote}{0}

\section{Spiky strings in the Bethe Ansatz:  $2$-cuts solution}

We start again with equation (\ref{ba}) where we do not rescale $u$ by $S$, i.e. we introduce the root distribution $\rho_0(u)=\sum_{k=1}^{S} \delta(u- u_k)$
\begin{eqnarray}
\ppint d u' \frac{\rho_0(u')}{u-u'} &=& \pi n_w(u)-\frac{J}{2 u}\equiv \pi N_w(u)  \label{htg}
\end{eqnarray}
The condition that the total momentum should vanish gives
\begin{equation}
\int d u \rho_0(u) \ln \frac{u+\frac{i}{2}}{u- \frac{i}{2}}=0  \label{mm1}
\end{equation}
The root density $\rho_0$ is normalized as
\begin{equation}
\int d u \rho_0(u)=S
\end{equation}
The $1$-loop anomalous dimension is
\begin{equation}
\frac{E-S-J}{2g^2} = \int d u \frac{\rho_0(u)}{u^2+ \frac{1}{4}}
\end{equation}

Let us consider two cuts $d \leq c \leq b \leq a$. Extending the bosonic wave number distribution from the previous section we take
\begin{eqnarray}
n_w(u)=\left\{
       \begin{array}{ll}
         -1, & d < u < c \\
         n-1, & b < u < a
       \end{array}
     \right. \label{alp1}
\end{eqnarray}

 Then the root distribution function is
\begin{eqnarray}
\rho_{0L} (u')&=&- \frac{2 n}{\pi} \sqrt{\frac{(c - u') (b - u')}{(u' - d)(a - u')}} \frac{1}{
  \sqrt{(a - c) (b - d)}} \bigg((a -
      d) \Pi[\frac{(b - a) (u' - d)}{(b - d)(u' - a)},
     r] + (u' - a) K[r]\bigg)\nonumber\\
&+&\frac{J}{2 \pi^2 u'}(B_1(u')-B_2(u'))  \label{rho1}
\end{eqnarray}
for the left interval $u'\in [d,c]$, and
\begin{eqnarray}
\rho_{0R}(u')&=& \frac{2n}{\pi} \sqrt{\frac{(u' - c) (u' - b)}{(u' - d)(a - u')}} \frac{1}{
  \sqrt{(a - c) (b - d)}} \bigg((d -
      a) \Pi[\frac{(d - c) (u'- a)}{(a - c) (u' - d)},
     r] + (u' - d) K[r]\bigg)\nonumber\\
&+&\frac{J}{2 \pi^2 u'}(-B_1(u')+B_2(u'))  \label{rho2}
\end{eqnarray}
for the right one $u'\in [b,a]$.
Here
\begin{equation}
r= \frac{(a - b) (c - d)}{(a - c)(b - d)}
\end{equation}
and we used the following definitions of the elliptic integrals
\begin{equation}
K[m]= \int_0^{\frac{\pi}{2}}\frac{d \alpha}{\sqrt{1- m \sin^2 \alpha}}, \quad \quad \Pi[n,m]=\int_0^{\frac{\pi}{2}}\frac{d \alpha}{\sqrt{1- m \sin^2 \alpha}}\frac{1}{1- n \sin^2 \alpha}
\end{equation}

Also, we denote $B_1,B_2$ the integrals
\begin{equation}
B_1(u') = \sqrt{|F(u')|} A_4, \quad \quad \quad B_2(u')=\sqrt{|F(u')|} A_5
\end{equation}

We compute the residue term in (\ref{fii}) and obtain
\begin{eqnarray}
 \pi \res{\frac{\Gg(w)}{w-z}}{\infty}
&=& z [n A_1 + \frac{J}{2 \pi}(A_4-A_5)] + A_2+ (n-1) A_3 \nonumber\\
&-& \frac{a+b+c+d}{2}[n A_1+ \frac{J}{2\pi}(A_4-A_5)]  \label{pil}
\end{eqnarray}
where $A_i$ integrals are defined in the Appendix A and can be written in terms of elliptic integrals.

As explained in Appendix A we need to take the above residue to zero in order for the root distribution (\ref{rho1},\ref{rho2}) to be a solution of (\ref{htg}). This gives the following relationship
\begin{equation}
n A_1 + \frac{J}{2\pi}(A_4-A_5)=0, \quad \quad \quad \quad \quad A_2+ (n-1) A_3=0  \label{eoi}
\end{equation}
Then one can write $J,n$ as functions of $a,b,c,d$
\begin{equation}
J= \frac{2 \pi n A_1}{A_5- A_4}, \quad \quad \quad n= 1- \frac{A_2}{A_3}  \label{eoi2}
\end{equation}

The normalization condition can be computed using the function $\Gg$ as
\begin{eqnarray}
\int \rho_0(u) d u &=& i \res{\Gg(w)}{\infty}\\
&=&  n A_1 + \frac{J}{2\pi}(A_4-A_5)+ A_2 +(n-1) A_3 + \frac{J}{2\pi}(A_2-A_3) + A_6 +(n-1) A_7  \nonumber
\end{eqnarray}
Taking into account equation (\ref{eoi}) the normalization condition reduces to
\begin{equation}
\frac{J}{2\pi}(A_2- A_3) + A_6 +(n-1) A_7=S  \label{norm}
\end{equation}

The consistency of equation (\ref{htg}) gives the equation
\begin{equation}
J \int d u \frac{\rho_0(u)}{u}= 2 \pi \int d u \rho_0(u) n(u)
\end{equation}
The momentum condition (\ref{mm1}) expanded in large $u$ gives
\begin{equation}
\int d u \frac{\rho_0(u)}{u}=0  \label{mm}
\end{equation}
We use again the residues of the function $\Gg$ to compute the integral in (\ref{mm})
\begin{equation}
\int d u \frac{\rho_0(u)}{u}=\frac{1}{2}\oint_{\mathcal{C}} d w \frac{\Gg(w)}{w}= i \pi  \res{\frac{\Gg(w)}{w}}{\infty}
   + i \pi \res{\frac{\Gg(w)}{w}}{0} \label{qr}
\end{equation}
We already computed the first term and set it to zero in (\ref{pil}). What remains is to compute the second term in (\ref{qr}).
We obtain
\begin{equation}
\int d u \frac{\rho(u)}{u}= \sqrt{a b c d}(-A_4- (n-1) A_5 -\frac{J}{2\pi}A_8 + \frac{J}{2\pi}A_9)
\end{equation}
Therefore the momentum condition (\ref{mm}) gives the condition
\begin{equation}
\sqrt{abcd} [A_4+ (n-1) A_5 + \frac{J}{2\pi}(A_8-A_9)]=0  \label{c2}
\end{equation}

All $A_i$ integrals depend on the parameters $a,b,c,d$. Let us write the physical charges in terms of these parameters
\begin{equation}
n= 1- \frac{A_2}{A_3}, \quad \quad  J= \frac{2 \pi A_1}{A_5-A_4}(1- \frac{A_2}{A_3})
\end{equation}
\begin{equation}
S= A_6+ \frac{A_1(A_2-A_3)^2}{A_3 (A_4-A_5)}-A_7 \frac{A_2}{A_3}, \quad \quad \frac{E - S-J}{2g^2}= 2 \pi \mathrm{Re}[\Gg(\frac{i}{2})]
\end{equation}
where we wrote the energy in terms of the function $G$ whose explicit expression in terms of elliptic integrals is in
(\ref{gii}).

We also have the condition (\ref{c2}) giving a relationship among parameters $a,b,c,d$
\begin{equation}
A_4 - \frac{A_2}{A_3}A_5 +\frac{ A_1 (A_8-A_9)(A_3-A_2)}{A_3 (A_5-A_4)}=0
\end{equation}
We observe that using the above equations one can in  principle solve for the unknown constants $a,b,c,d$
and obtain the $1$-loop anomalous dimension $E=E(n,S,J)$. However, this is not possible to find explicitly because
of the complicated elliptic integrals involved. Below we consider particular limits.

\subsection{Particular limit: 1 cut solution}

From the above general $2$-cut solution for an arbitrary number of spikes let us recover the $1$-cut solution again with arbitrary number of spikes. The $1$-cut solution should be obtained in the limit
\begin{equation}
b=c=0, \quad \quad d<0, \quad \quad a>0
\end{equation}
To obtain this limit we take $b=\epsilon$, $c=-\eta \epsilon$ and take $\epsilon \rightarrow 0_{+}$. The parameter $\eta>0$ is to be determined. The integrals in this limit become
\begin{equation}
A_1= - \frac{\ln \epsilon}{\sqrt{- a d}}+O(\epsilon^0), \quad \quad
A_2= -\frac{\pi}{2}+ \arcsin \frac{a+d}{a-d}, \quad \quad \quad A_3= \frac{\pi}{2}+ \arcsin \frac{a+d}{a-d}
\end{equation}
\begin{equation}
A_4= - \frac{2 \cot^{-1} \sqrt{\eta}}{\sqrt{-ad} \sqrt{\eta}}\frac{1}{\epsilon} +O(\ln \epsilon), \quad \quad
A_5= \frac{2 \tan^{-1} \sqrt{\eta}}{\sqrt{-ad} \sqrt{\eta}}\frac{1}{\epsilon} +O(\ln \epsilon)
\end{equation}
\begin{equation}
A_6=\sqrt{-a d }-\frac{\pi}{4}(a+d) -\frac{(a+d)}{2}\arcsin \frac{a+d}{d-a}
\end{equation}
\begin{equation}
A_7=\sqrt{-a d }+\frac{\pi}{4}(a+d) -\frac{(a+d)}{2}\arcsin \frac{a+d}{d-a}
\end{equation}
\begin{equation}
A_8=\frac{\sqrt{\eta}-(\eta-1) \cot^{-1}\sqrt{\eta}}{\sqrt{-a d}\eta \sqrt{\eta}}\frac{1}{\epsilon^2}+O(\ln \epsilon), \quad \quad
A_9= \frac{\sqrt{\eta}+(\eta-1) \tan^{-1}\sqrt{\eta}}{\sqrt{-a d}\eta \sqrt{\eta}}\frac{1}{\epsilon^2}+O(\ln \epsilon)
\end{equation}

Then the second equation in (\ref{eoi}) implies
\begin{equation}
\frac{a+d}{a-d}= - \cos \frac{\pi}{n} \label{par1}
\end{equation}
which indeed matches the result from the section 3.  The first equation in (\ref{eoi2}) becomes
\begin{equation}
J = -  \pi n \epsilon \ln \epsilon \frac{\sqrt{\eta}}{\tan^{-1} \sqrt{\eta}+\cot^{-1}\sqrt{\eta}}\rightarrow  0
\end{equation}
The normalization equation (\ref{norm}) implies
\begin{equation}
S= - \frac{n}{2}(a+d) \tan \frac{\pi}{n} \label{par2}
\end{equation}
which again matches the results in section 3.
Finally, the momentum condition (\ref{c2}) determines $\eta$ to be
\begin{equation}
\eta= 1+ \pi \sqrt{\frac{n-2}{n}}\frac{1}{\sqrt{\ln \epsilon}}+...
\end{equation}

Equations (\ref{par1}, \ref{par2}) can be used to find the parameters $a,d$ in terms of $S,n$. The Bethe root density (\ref{rho1}, \ref{rho2}) in this limit reduces precisely to the $1$-cut density expression (\ref{qma}).

\subsection{Particular limit: ``$pp$-wave'' type scaling}

We want to consider the limit of large $n$. To achieve this we take the limit $d\rightarrow - \infty$ while keeping $a,b,c$ fixed. The integrals become
\begin{equation}
A_1= \frac{2}{\sqrt{a-c}}K[q] \frac{1}{\sqrt{-d}}+..., \quad \quad q=\frac{a-b}{a-c}
\end{equation}
\begin{equation}
A_2= -\pi +O(\frac{1}{\sqrt{-d}}), \quad \quad
A_3= \frac{2}{\sqrt{a-c}}[c K[q]+(a-c) E[q]]\frac{1}{\sqrt{-d}}+...
\end{equation}
\begin{equation}
A_4= \frac{2}{a \sqrt{a-c}}[K[q]-\Pi[\frac{a}{a-c},q]] \frac{1}{\sqrt{-d}}+O(\frac{1}{(-d)^{3/2}})
\end{equation}
\begin{equation}
A_5= \frac{2}{a\sqrt{a-c}}\Pi[\frac{a-b}{a},q]\frac{1}{\sqrt{-d}}+..., \quad \quad
A_6= - \frac{\pi}{2}d - \frac{\pi}{2}(a+b+c) +O(\frac{1}{\sqrt{-d}})
\end{equation}
\begin{equation}
A_7= \frac{2}{3 \sqrt{a-c}}[2 (a-c) (a+b+c) E[q]+(a(c-b)+c(b+2 c))K[q]]\frac{1}{\sqrt{-d}}+...
\end{equation}
\begin{equation}
A_8=\frac{1}{a^2 b c \sqrt{a-c}}[a (c-a) E[q] + b(a+c) K[q]-(b c +a b + a c) \Pi[\frac{a}{a-c},q]]\frac{1}{\sqrt{-d}}+...
\end{equation}
\begin{equation}
A_9= \frac{1}{a^2 b c \sqrt{a-c}}[ a(c-a) E[q]- a c K[q]+(bc+ab +ac)\Pi[\frac{a-b}{a},q]]\frac{1}{\sqrt{-d}}+...
\end{equation}
The root distributions (\ref{rho1},\ref{rho2}) in the large $-d $ limit reduce to
\begin{equation}
\rho_{0L}(u')=-\frac{2 n}{\pi \sqrt{a-c}}\sqrt{\frac{(u'-b)(u'-c)}{a-u'}}\bigg(\frac{a-u'}{u'}K[q]+\Pi[\frac{a-b}{a-u'},q]\bigg)+O(\frac{1}{\sqrt{-d}})
\end{equation}
\begin{equation}
\rho_{0R}(u')=\frac{2 n}{\pi \sqrt{a-c}}\sqrt{\frac{(u'-b)(u'-c)}{a-u'}}\bigg(\frac{a}{u'}K[q]-\Pi[\frac{a-u'}{a-c},q]\bigg)+O(\frac{1}{\sqrt{-d}})
\end{equation}

The number of spikes in this limit is
\begin{equation}
n=\frac{\pi}{2}\frac{\sqrt{a-c}}{c K[q]+(a-c)E[q]}\sqrt{-d}+...
\end{equation}
In this limit the physical parameters scale as
\begin{equation}
n \gg 1, \quad \quad S \sim n^2, \quad \quad J \sim n,\quad \quad E-S \sim n
\end{equation}
which is the same as the $pp$-wave limit scaling found at strong coupling in \cite{kt,iktt}.
More precisely we obtain
\begin{equation}
S= \frac{2}{\pi}\frac{(c K[q]+(a-c) E[q])^2}{a-c} n^2, \quad \quad
J= \frac{2 \pi a K[q]}{\Pi[\frac{a-b}{a},q]+\Pi[\frac{a}{a-c},q]-K[q]}n  \label{gad1}
\end{equation}
The condition (\ref{c2}) gives the relationship
\begin{equation}
2 \Pi[\frac{a-b}{a},q]= \frac{ab + bc + ac }{bc} K[q]  \label{gad2}
\end{equation}
The above three equations are to be solved for $a,b,c$ as functions of $S,J,n$ and then plug them in the energy.

In the large $d$ limit the function $\Gg(w)$ becomes
\begin{equation}
\Gg(w)=- \frac{i \sqrt{(w-a)(w-b)(w-c)}}{w (w-a)}
\frac{H(w) \sqrt{-d}}{((a-c) E[q]+c K[q])(K[q]-\Pi[\frac{a-b}{a},q]-\Pi[\frac{a}{a-c},q])}+...
\end{equation}
where
\begin{eqnarray}
H(w)&=&w K[q]^2 + w \big(\Pi[\frac{a-b}{a},q]+\Pi[\frac{a}{a-c},q]\big)\Pi[\frac{a-b}{a-w},q]\\
&+& K[q]\big[(a-w) \Pi[\frac{a-b}{a},q]+(a-w)\Pi[\frac{a}{a-c},q]-(a+w) \Pi[\frac{a-b}{a-w},q]-a \Pi[\frac{a-w}{a-c},q]\big]\nonumber
\end{eqnarray}
To find the energy we need to take $w=\frac{i}{2}$ in $\Gg(w)$. The end of the right cut $a$ is still large, i.e. $|d| \gg a \gg 1 $, therefore we can expand in large $a$ and finally obtain the energy at leading order in large $n$
\begin{equation}
\frac{E-S-J}{2g^2}= \frac{2 n }{\sqrt{a (a-c)}} (1+4 b^2)^{\frac{1}{4}}(1+4 c^2)^{\frac{1}{4}} \sin[\frac{1}{2}(\tan^{-1}\frac{1}{2b}+\tan^{-1}\frac{1}{2c})] (K[q]-\Pi[\frac{a-b}{a},q]) \label{dja}
\end{equation}
When $b, c$ are also large, i.e $|d| \gg b, c  \gg 1$ the expression simplifies
\begin{equation}
\frac{E-S-J}{2g^2}= - \frac{n (b+c)}{\sqrt{a-c}\sqrt{-a b c}}(K[q]-\Pi[\frac{a-b}{a},q])  \label{dja1}
\end{equation}
We therefore obtain a set of three equations (\ref{gad1}, \ref{gad2}) which are to be solved for $a,b,c$ in terms of $S,J,n$. Furthermore, such solution for $a,b,c$ should be replaced in the energy (\ref{dja}, \ref{dja1}).

\bigskip

\section{Conclusions}

 We have found the Bethe root distribution that describes the spiky strings in the context of the spin chain model that describes the
field theory operators in the $SL(2)$ sector. At one-loop we constructed solutions where the roots condense on one cut and two cuts.
The one cut solutions we were able to extend to all-loops. In particular expanding the result for large coupling we see that at order $\ln S$ and $S^0$
in the large $S$ expansion, we recover exactly the result obtained from the classical solutions. The result contains a non-trivial dependence on the number of spikes and therefore is an important test of both,
our proposal of describing the spiky strings by this particular Bethe
ansatz solution and, more broadly, of the
all-loop Bethe ansatz used in the calculation.
 In fact, the all-loop Bethe ansatz also provides a prediction for the
one-loop quantum corrections to the spiky string. This is a doable calculation that should be interesting to perform in order to verify the prediction.
 In the case of the two cut solution, we did not extend the solution to all-loops. This should be interesting further work since the result at strong
coupling is known from the string side. We have found that the equations simplify in a pp-wave-like limit where the number of spikes $n$ grows to infinity keeping $\frac{E+S}{n^2}$, $\frac{E-S}{n}$ and $\frac{J}{n}$ fixed. We showed that such a limit is well defined at one-loop suggesting that it actually can be taken also at higher loops. Finding the all-loop solution in this limit would be an interesting problem for further studies.

\section*{Acknowledgments }

We are  grateful to N. Gromov,  J. Minahan, A. Tseytlin and S. Zieme  for useful  discussions.
 M. K. and A.T. were supported in part by NSF under grant PHY-0805948 and DOE under grant DE-FG02-91ER40681. L.F. was supported in part by the Swedish research council.

\bigskip

\appendix
\subsection*{Appendix A:  Solving the Bethe equations}

\refstepcounter{section}
\def\theequation{A.\arabic{equation}}
\setcounter{equation}{0}

 In this appendix we summarize the procedure that we used to solve the Bethe Ansatz equations. Although such methods are well known we
adapt them here to our particular needs. However, it remains generic enough to be applied to other situations (e.g. at strong coupling.).

 To be concrete we have to find a function $\rho(x)$ defined on the union of several segments $C_{i=1\ldots n}$ on the real axis (\ie\ cuts) such that
\beq
\frac{1}{\pi} \ppint_{C_1\cup \ldots C_n} \frac{\rho(x')}{x'-x} = N(x) , \ \ \ \ x\in C_1\cup \ldots  C_n
\label{eqtosolve}
\eeq
where $\ppint$ denotes principal part of the integral and $N(x)$ is a given real function defined on the cuts. As a first step we have to find
a complex function $\Gg(z)$, analytic except for cuts at $C_i$ and a pole at infinity. Furthermore, we require that, on the cuts, the real
part changes sign and the imaginary part is equal to $N(x)$:
\beq
 \Gg(x\pm i\epsilon) = \pm \tilde{\rho}(x) + i N(x), \ \ \ \ x\in C_1\cup \ldots C_n , \ \ \ \epsilon\rightarrow 0^+.
\label{jumps1}
\eeq
The, yet undetermined, real part we denote as $\tilde{\rho}(x)$. It is straightforward to find such a function in general for any $N(x)$. While the procedure following below can also be directly extended for an arbitrary number of cuts, we take for simplicity two cuts.

\begin{figure}
\begin{center}
\epsfig{file=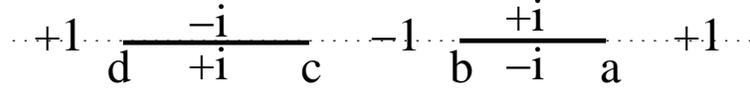, width=10cm}
\end{center}
\caption{Phase of the function $F(w)$ defined in the text, when $w$ approaches the real axes. Along the two cuts, the imaginary part
changes sign as depicted. }
\label{fig:cuts1}
\end{figure}

\begin{figure}
\begin{center}
\epsfig{file=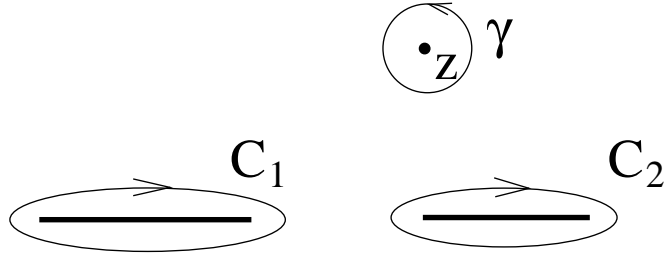, width=10cm}
\end{center}
\caption{Contours of integration surrounding $z$ and the two cuts.
 }
\label{fig:cuts2}
\end{figure}

To be specific lets consider two cuts
$C_1=[d,c]$, $C_2=[b,a]$ with $d<c<b<a$ and the following function:
\beq
 F(w) = \sqrt{w-a}\sqrt{w-b}\sqrt{w-c}\sqrt{w-d}
\eeq
where the square roots are defined with a standard cut on the negative real axis. In figure \ref{fig:cuts1} we show the phase of $F(w)$ for
$w$ close to the real axis. Now we can write
\beq
 \Gg(w) = \frac{1}{\pi} \int_{C_1\cup C_2} \frac{F(w)}{F(x+i\epsilon)} \frac{N(x)}{x-w} dx
\eeq
It is easy to verify that $\Gg(w)$ so defined has cuts in $C_1$ and $C_2$ and satisfies eq.(\ref{jumps1}) with
\beq
\tilde{\rho}(x) = \frac{1}{\pi} \ppint \left|\frac{F(x)}{F(x')}\right|\frac{N(x')}{x'-x} dx'
\label{rhosol}
\eeq
Furthermore it behaves as $\Gg(w)\sim w (\w\rightarrow\infty)$. Now consider an arbitrary point $z$ (away from the cuts) and a small contour
$\gamma$ encircling it as shown in figure \ref{fig:cuts2}. We have
\beq
 \Gg(z) = \frac{1}{2\pi i} \oint_\gamma \frac{\Gg(w)}{w-z} dw
\eeq
Deforming the contour we obtain
\beq
 \Gg(z) = -\res{\frac{\Gg(w)}{w-z}}{\infty} + \frac{1}{2\pi i} \oint_{C_1\cup C_2} \frac{\Gg(w)}{w-z} dw
\eeq
 where the integral on the right hand side is over contours encircling the cuts clockwise. From the property (\ref{jumps1}) we find
\beq
 \Gg(z) = -\res{\frac{\Gg(w)}{w-z}}{\infty} + \frac{1}{\pi i} \int_{C_1\cup C_2} \frac{\rho(x')}{x'-z} dx'
\eeq
 From this result and (\ref{jumps1}) we find that, if $z$ approaches one of the cuts from above, then:
\beqa
\tilde{\rho}(x) + i N(x) &=& \Gg(x+i\epsilon) = -\res{\frac{\Gg(w)}{w-x}}{\infty}
                      + \frac{1}{\pi i} \int_{C_1\cup C_2} \frac{\tilde{\rho}(x')}{x'-x-i\epsilon} dx' \\
                 &=&  -\res{\frac{\Gg(w)}{w-x}}{\infty} +\tilde{\rho}(x)
                       + \frac{1}{\pi i} \ppint_{C_1\cup C_2} \frac{\tilde{\rho}(x')}{x'-x} dx',
                       \ \ \ \epsilon\rightarrow 0^+
\eeqa

Therefore
\beq
N(x) = i \res{\frac{\Gg(w)}{w-x}}{\infty} + \frac{1}{\pi} \ppint_{C_1\cup C_2} \frac{\tilde{\rho}(x')}{x-x'} dx',
\eeq
So we obtain that $\tilde{\rho}(x)$ solves the problem (\ref{eqtosolve}) under the condition
\beq
 \res{\frac{\Gg(w)}{w-x}}{\infty} = 0  \label{fii}
\eeq
The residue at infinity can be computed by expanding
\beq
 \Gg(w) = \Gg_0 w+ \Gg_1 + \frac{1}{\w} \Gg_2 +\ldots
\eeq
which results in
\beq
\res{\frac{\Gg(w)}{w-x}}{\infty} = - \Gg_0 - x \Gg_1
\eeq
namely we need $\Gg_0=\Gg_1=0$. Since $\Gg(z)$ is already fixed we can only solve this problem for certain positions of the cuts.
Namely, $\Gg_0=\Gg_1=0$ is an equation for $a$,$b$,$c$,$d$. In that case, formula (\ref{rhosol}) gives the solution to the problem (\ref{eqtosolve}).

Furthermore, it is now straight forward to compute integrals of the type
\beq
 \int_{C_1\cup C_2} \frac{\rho(u)}{u^n} du = \half \oint_{C_1\cup C_2} \frac{\Gg(w)}{w^n} dw = \pi i \res{\frac{\Gg(w)}{w^n}}{0}
+\pi i \res{\frac{\Gg(w)}{w^n}}{\infty}
\eeq
For example
\beqa
 \int_{C_1\cup C_2} \rho(u) du & =& \half \oint_{C_1\cup C_2} \Gg(w) dw = \pi i \res{\Gg(w)}{\infty} = - i\pi \Gg_2 \\
 \int_{C_1\cup C_2} \frac{\rho(u)}{u} du &=& \half \oint_{C_1\cup C_2} \frac{\Gg(w)}{w} dw = \pi i \Gg(0) -\pi i \Gg_0 \\
 \int_{C_1\cup C_2} \frac{\rho(u)}{u-z} du &=& \half \oint_{C_1\cup C_2} \frac{\Gg(w)}{w-z} dw = i \pi \Gg(z)
\eeqa
and so on. In particular the energy $E$ and spin $S$ can be computed by these formulas. Also, the last equation can be used as an alternative definition of $\Gg(z)$ \cite{es}. In particular we can recover (\ref{jumps1}) by noticing that
\beqa
i \pi G(u' \pm i \epsilon) &=&  \int_{C_1\cup C_2} \frac{\rho(u)}{u-u' \mp i \epsilon} du=  \ppint_{C_1\cup C_2} \frac{\rho(u)}{u-u'} du \pm i \pi
 \rho(u')\nonumber\\
 &=& - \pi N(u') \pm i \pi \rho(u')
\eeqa

\bigskip

The above discussion of two cuts is valid for any function $N(u)$. However, in this paper $N(u)$ is simply (\ref{htg}) $N(u)=n_w(u)-\frac{J}{2 \pi u}$. For such $N(u)$, the function $\Gg$ can be written in terms of elliptic integrals as
\begin{eqnarray}
\Gg(w)&=& \frac{2 i \sqrt{(w-a)(w-b)(w-c)(w-d)}}{\pi (w-a)(w-d) \sqrt{(a-c)(b-d)}}\bigg[(n_2-\frac{J}{2 \pi w})\nonumber\\
&\times & \bigg((a-d)\Pi[\frac{(b-a)(w-d)}{(b-d)(w-a)},r]+
(w-a) K[r]\bigg)
-(n_1-\frac{J}{2 \pi w})\nonumber\\
&\times& \bigg((d-a) \Pi[\frac{(d-c)(w-a)}{(a-c)(w-d)},r]+(w-d) K[r]\bigg)\bigg]\nonumber\\
&+&\frac{i J \sqrt{(w-a)(w-b)(w-c)(w-d)}}{2 \pi^2 w}(A_4-A_5)  \label{gii}
\end{eqnarray}
where $n_1,n_2$ are integers. In our case of interest $n_1=-1$, $n_2=n-1$.

\bigskip

\bigskip

In the remaining of this appendix let us define the integrals used in this paper. First it is convenient to define the function
\beq
|F(x)| = \sqrt{|(x-a)(x-b)(x-c)(x-d)|}
\eeq
We can now compute
\begin{equation}
A_1= \int_b^a \frac{d x}{|F(x)|}= \frac{2}{\sqrt{(a-c)(b-d)}}K[r]
\end{equation}
\begin{equation}
A_2= \int_d^c \frac{x dx}{|F(x)|}=\frac{2}{\sqrt{(a-c)(b-d)}}[(d-a) \Pi[\frac{d-c}{a-c},r]+a K[r]]
\end{equation}
\begin{equation}
A_3= \int_b^a \frac{x dx}{|F(x)|}=\frac{2}{\sqrt{(a-c)(b-d)}}[(a-d) \Pi[\frac{b-a}{b-d},r]+d K[r]]
\end{equation}
\begin{equation}
A_4= \int_d^c \frac{dx}{x |F(x)|}=\frac{2}{ad \sqrt{(a-c)(b-d)}}[(a-d) \Pi[\frac{a(d-c)}{d(a-c)},r]+d K[r]]
\end{equation}
\begin{equation}
A_5= \int_b^a \frac{dx}{x |F(x)|}=\frac{2}{ad \sqrt{(a-c)(b-d)}}[(d-a) \Pi[\frac{d(b-a)}{a(b-d)},r]+a K[r]]
\end{equation}
\begin{eqnarray}
A_6&=& \int_d^c \frac{x^2 dx }{ |F(x)|}=\frac{1}{\sqrt{(a-c)(b-d)}}[(a-c)(b-d) E[r]\nonumber\\
&+&(a(a+c)+d(a-c))K[r]-(a-d)(a+b+c+d)\Pi[\frac{d-c}{a-c},r]]
\end{eqnarray}
\begin{eqnarray}
A_7&=& \int_b^a \frac{x^2 dx}{ |F(x)|}=\frac{1}{\sqrt{(a-c)(b-d)}}[(a-c)(b-d) E[r]\nonumber\\
&+&(c(a+c)+b(c-a))K[r]-(c-b)(a+b+c+d)\Pi[\frac{a-b}{a-c},r]]
\end{eqnarray}
\begin{eqnarray}
A_8&=& \int_d^c \frac{dx}{x^2 |F(x)|}=\frac{1}{a^2 d^2 b c \sqrt{(a-c)(b-d)}}[a d (a-c)(b-d) E[r]\\
&+& b d (-a^2 +c d +a (c+d)) K[r]+(a-d)(bcd+acd +abc +a b d )\Pi[\frac{a (d-c)}{d (a-c)},r]]\nonumber
\end{eqnarray}
\begin{eqnarray}
A_9&=& \int_b^a \frac{dx}{x^2 |F(x)|}=\frac{1}{c^2 b^2 a d \sqrt{(a-c)(b-d)}}[c b (a-c)(b-d) E[r]\\
&+& b d (-c^2 +a b  +c (a+b)) K[r]+(c-b)( d a b+ c a b +a c d + c b d )\Pi[\frac{c (b-a)}{b (c-a)},r]]\nonumber
\end{eqnarray}


\appendix
\subsection*{Appendix B: Subleading corrections for the ground state }

\refstepcounter{section}
\def\theequation{B.\arabic{equation}}
\setcounter{equation}{0}

In the main text we compute the subleading corrections (in the large $S$ expansion) to the density of roots. It is convenient to remind ourselves how
this is done in the case of the ground state which is described by a symmetric root distribution. We do this at one-loop and and then make us of that solution to construct the all-loop solution.
\subsubsection*{One loop}
The one-loop equation for the ground state reads
\begin{equation}\label{one-loopground}
2\pi\rho_0(u)+2\pi(J-2)\delta(u)-\frac{J}{u^2+1/4}-2\int_{-S/2}^{S/2}dv\frac{\rho_0(v)}{(u-v)^2+1}=0.
\end{equation}
An approximate solution to the equation is given by the following density \cite{K} \cite{es} which we expand for large $S$
\begin{equation}
\tilde{\rho}_0(u)=\frac{1}{\pi}\ln\frac{1+\sqrt{1-4u^2/S^2}}{1-\sqrt{1-4u^2/S^2}}=\frac{2}{\pi}\ln S-\frac{1}{\pi }\ln(u^2)+\mathcal{O}(1/S^2).
\end{equation}
This density captures the leading order in the large spin expansion correctly. To get the subleading corrections of order $S^0$ we split the density,
\begin{equation}
\rho_0(u)=\tilde{\rho}_0(u)+r(u),
\end{equation}
and solve for $r(u)$. We are now allowed to extend the limits of integration to $\pm\infty$.
\begin{eqnarray}
&&2\pi r(u)+4\ln S-2\ln(u^2)+2\pi(J-2)\delta(u)-\frac{J}{u^2+1/4}\nonumber\\
&&-2\int_{-\infty}^\infty dv\left(\frac{2}{\pi}\ln S-\frac{1}{\pi }\ln(v^2)\right)\frac{1}{(u-v)^2+1}-2\int_{-\infty}^\infty dv\frac{r(v)}{(u-v)^2+1}.
\end{eqnarray}
The equation can now be solved by Fourier transformation
\begin{equation}
\hat{r}(t)=2\left(\frac{e^{-|t|/2}}{1-e^{-|t|}}-\frac{1}{|t|}\right)-(J-2)\frac{1}{1+e^{-|t|/2}}.
\end{equation}
Computing the energy we find
\begin{eqnarray}\label{one-loopEnergyA}
\nonumber \frac{E-S-J}{2g^2}&=&\int_{-S/2}^{S/2}\frac{\tilde{\rho}_0(u)}{u^2+1/4}+\int_{-\infty}^\infty du\frac{r(u)}{u^2+1/4}\\
\nonumber&=&4\ln S+4\ln 2+2\int_{-\infty}^\infty dt\left(\frac{e^{-|t|}}{1-e^{-|t|}}-\frac{e^{-|t|/2}}{|t|}\right)-(J-2)\int_{-\infty}^\infty dt\frac{e^{-|t|/2}}{1+e^{-|t|/2}}\\
&=&4\left(\ln S+\gamma_E-(J-2)\ln 2\right)
\end{eqnarray}

\subsubsection*{All loops}
The all loop equation reads
\begin{eqnarray}
&&J\frac{d}{du}\ln\frac{x(i/2+u)}{x(i/2-u)}=2\pi\rho(u)+2\pi(J-2)\delta(u)-\int_{-S/2}^{S/2}\frac{\rho(u')}{(u-u')^2+1}\nonumber\\
&&+\int_{-S/2}^{S/2}du'\rho(u')\frac{d}{du}\left(2i\ln\frac{1-g^2/x^+(u)x^-(u')}{1-g^2/x^+(u')x^-(u)}-2\theta(u,u')\right).
\end{eqnarray}
In the following the dressing phase will be omitted for simplicity. It is easily restored in the end.
Splitting the density
\begin{equation}
 \rho(u)=\rho_0(u)+2g^2\tilde{\sigma}(u),
 \end{equation}
 where $\rho_0(u)$ satisfies \eqref{one-loopground}
we find
\begin{eqnarray}\label{all-loopBEA}
&&\frac{J}{2g^2}\left(\frac{d}{du}\ln\frac{x(i/2+u)}{x(i/2-u)}-\frac{1}{u^2+1/4}\right)=2\pi\tilde{\sigma}-2\int_{-\infty}^\infty\frac{\tilde{\sigma}(u')}{(u-u')^2+1}+I(u)\nonumber\\
&&+\int_{-\infty}^{\infty}du'\tilde{\sigma}(u')\frac{d}{du}\left(2i\ln\frac{1-g^2/x^+(u)x^-(u')}{1-g^2/x^+(u')x^-(u)}\right)
\end{eqnarray}
where
\begin{eqnarray}
I(u)=\frac{1}{2g^2}\int_{-S/2}^{S/2}du'\rho_0(u')\frac{d}{du}\left(2i\ln\frac{1-g^2/x^+(u)x^-(u')}{1-g^2/x^+(u')x^-(u)}\right).
\end{eqnarray}
Note that the limits of integration in the integrals containing the higher loop density can extended to $\pm\infty$ as we are considering the large $S$ limit. Further we can split the integral, $I(u)$, into two parts
\begin{eqnarray}\label{I(u)A}
\nonumber I(u)&=&\pi  \left(G(\tfrac{i}{2})-G(-\tfrac{i}{2})\right)\frac{d}{du}\left(\frac{1}{x^+(u)}-\frac{1}{x^-(u)}\right)\\
&+&\int_{-\infty}^\infty du'\rho_0(u')\left[\frac{1}{2g^2}\frac{d}{du}\left(2i\ln\frac{1-g^2/x^+(u)x^-(u')}{1-g^2/x^+(u')x^-(u)}\right)\right.\nonumber \\
&-&\left.\frac{1}{u'^2+1/4}\frac{d}{du}\left(\frac{1}{x^+(u)}-\frac{1}{x^-(u)}\right)\right]
\end{eqnarray}
where $G(u)$ is the one-loop resolvent, defined similarly to \eqref{defresolvent}.  Here we have used that the root distribution is symmetric around the origin. Note that in the integral in \eqref{I(u)A} we have extended the integration limits to $\pm \infty$.

Fourier transformation of the equation \eqref{all-loopBEA} gives
\begin{eqnarray}
\nonumber \frac{J}{2g^2}2\pi e^{-t/2}\left(J_0(2gt)-1\right)&=&2\pi\hat{\tilde{\sigma}}(t)(1-e^{-t})+\hat{I}(t)\\
&+&8\pi g^2te^{-t/2}\int_0^\infty dt'\hat{\tilde{\sigma}}(t')e^{-t'/2}K(2gt,2gt'),
\end{eqnarray}
with the kernel given by \eqref{kernelwithoutdressing}.
For $\hat{I}(t)$ we find, to first two orders in the expansion,
\begin{eqnarray}
\lefteqn{\hat{I}(t)=16\left(\ln S+\gamma_E-(J-2)\ln 2\right)\pi t e^{-t/2}K(2gt,0)}\\
&+&4\pi t e^{-t/2}\int_0^\infty dt' \left(\frac{2}{1-e^{-t'}}-\frac{J-2}{1+e^{-t'/2}}\right)e^{-t'/2}\left(K(2gt,2gt')-K(2gt,0)\right)\nonumber.
\end{eqnarray}
Redefining the density as
\begin{equation}
\hat{\sigma}(t)=-\frac{1}{4}e^{-t/2}\hat{\tilde{\sigma}}(t)
\end{equation}
we hence find for the all-loop equation
\begin{eqnarray}\label{eq: complete densityA}
\hat{\sigma}(t)&=&\frac{t}{e^t-1} \Big[ K(2gt,0)(\ln S +\gamma_E-(J-2)\ln2) \nonumber-\frac{J}{8g^2 t}(J_0(2gt) - 1)\\
\nonumber& +& \frac{1}{2}\int_{0}^{\infty} dt' \Big( \frac{2}{e^{t'}-1}-\frac{J-2}{e^{t'/2}+1}\Big)\left(K(2gt,2gt')-K(2gt,0)\right) \\
&-& 4g^2 \int_{0}^{\infty} dt' K(2gt,2gt')\hat{\sigma}(t') \Big] \,.
\end{eqnarray}
The energy is then given by
\begin{equation}
E-S=J+16g^2\hat{\sigma}(0).
\end{equation}
The effect of the dressing phase can now be restored by replacing the kernel above with \eqref{kernelwithdressing} and we have hence obtained the same integral equation as in \cite{fz}.

Note that this derivation explicitly made use of the fact that the density is symmetric. In the main text we do a similar calculation for the
asymmetric distribution we are considering in this paper.

\end{document}